\documentclass[compsoc, conference, a4paper, 10pt, times]{IEEEtran}

\usepackage{cite}
\usepackage{amsmath,amssymb,amsfonts}
\usepackage{graphicx}
\usepackage{textcomp}
\usepackage[svgnames]{xcolor}
\def\BibTeX{{\rm B\kern-.05em{\sc i\kern-.025em b}\kern-.08em
    T\kern-.1667em\lower.7ex\hbox{E}\kern-.125emX}}

\usepackage{algorithm}
\usepackage{algpseudocode}
\usepackage{tikz}

\newcommand{\bheading}[1]{{\vspace{2pt}\noindent{\textbf{#1}}\hspace{2pt}}}

\newcommand{\fullcirc}{\raisebox{-0.5mm}{\includegraphics[scale=0.025]{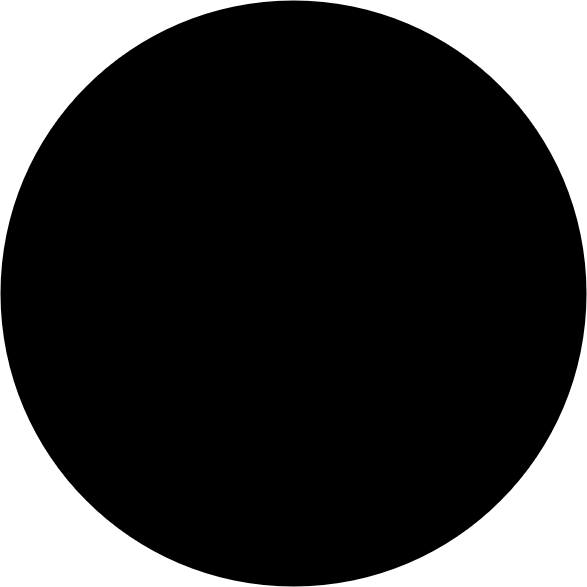}}}
\newcommand{\voice}{\raisebox{-0.3mm}{\includegraphics[scale=0.03]{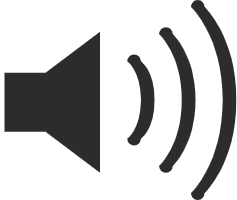}}}
\newcommand{\wave}{\raisebox{-0.3mm}{\includegraphics[scale=0.03]{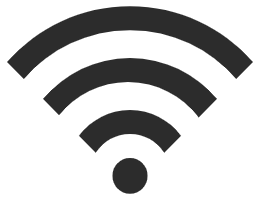}}}
\newcommand{\laser}{\raisebox{-0.8mm}{\includegraphics[scale=0.03]{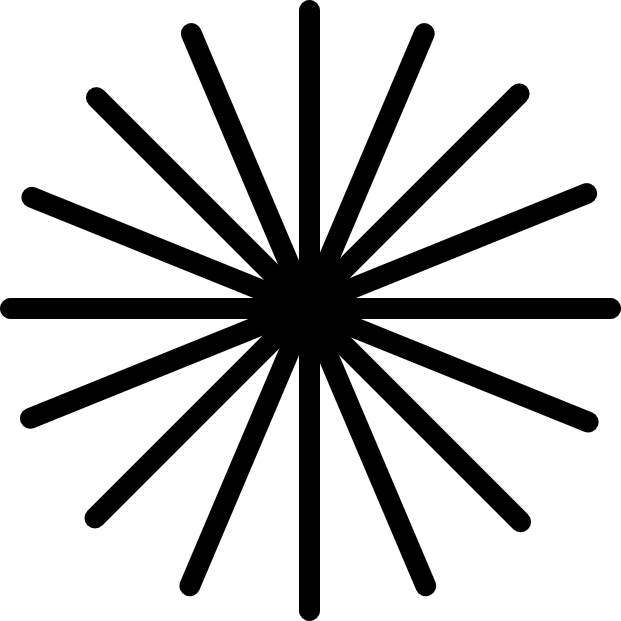}}}
\newcommand{\gps}{\raisebox{-0.6mm}{\includegraphics[scale=0.04]{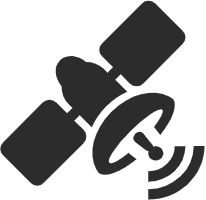}}}
\newcommand{\shape}{\raisebox{-0.6mm}{\includegraphics[scale=0.04]{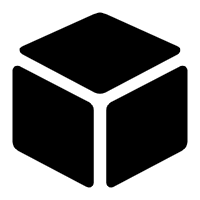}}}
\newcommand{\sticker}{\raisebox{-0.6mm}{\includegraphics[scale=0.04]{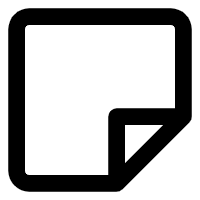}}}
\newcommand{\car}{\raisebox{-0.4mm}{\includegraphics[scale=0.06]{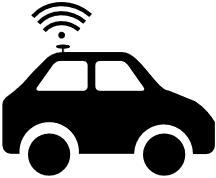}}}
\newcommand{\drone}{\raisebox{-0.2mm}{\includegraphics[scale=0.035]{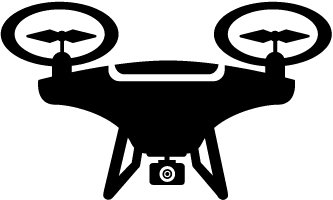}}}
\newcommand{\mb}{\raisebox{-0.4mm}{\includegraphics[scale=0.06]{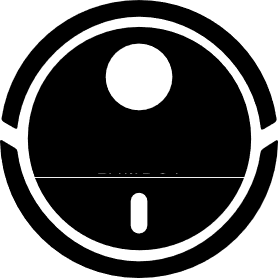}}}
\newcommand{\indoor}{\raisebox{-1.0mm}{\includegraphics[scale=0.05]{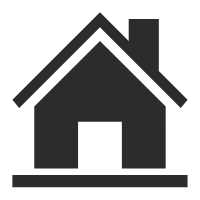}}}
\newcommand{\outdoor}{\raisebox{-0.7mm}{\includegraphics[scale=0.045]{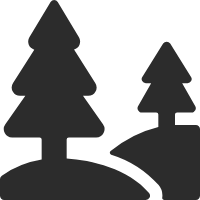}}}

\newcommand{\ratk}{\raisebox{-0.7mm}{\includegraphics[scale=0.050]{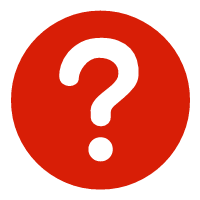}}}
\newcommand{\batk}{\raisebox{-0.7mm}{\includegraphics[scale=0.050]{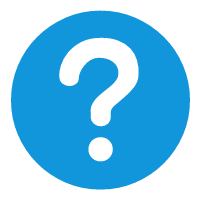}}}
\newcommand{\gatk}{\raisebox{-0.7mm}{\includegraphics[scale=0.050]{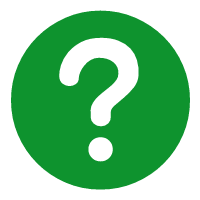}}}
\newcommand{\one}{\raisebox{-0.6mm}{\large{\ding{182}}}}
\newcommand{\two}{\raisebox{-0.6mm}{\large{\ding{183}}}}
\newcommand{\three}{\raisebox{-0.6mm}{\large{\ding{184}}}}
\newcommand{\four}{\raisebox{-0.6mm}{\large{\ding{185}}}}
\newcommand{\five}{\raisebox{-0.6mm}{\large{\ding{186}}}}
\newcommand{\six}{\raisebox{-0.6mm}{\large{\ding{187}}}}
\newcommand{\seven}{\raisebox{-0.6mm}{\large{\ding{188}}}}
\newcommand{\eight}{\raisebox{-0.6mm}{\large{\ding{189}}}}
\newcommand{\nine}{\raisebox{-0.6mm}{\large{\ding{190}}}}
\newcommand{\ten}{\raisebox{-0.6mm}{\large{\ding{191}}}}
\newcommand{\eleven}{\raisebox{-1.2mm}{\includegraphics[scale=0.50]{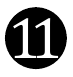}}}

\newcommand{\arone}{\raisebox{-0.7mm}{\includegraphics[scale=0.50]{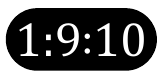}}}
\newcommand{\artwo}{\raisebox{-0.7mm}{\includegraphics[scale=0.50]{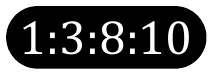}}}
\newcommand{\arthree}{\raisebox{-0.7mm}{\includegraphics[scale=0.50]{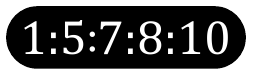}}}
\newcommand{\arfour}{\raisebox{-0.7mm}{\includegraphics[scale=0.50]{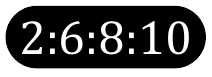}}}
\newcommand{\arfive}{\raisebox{-0.7mm}{\includegraphics[scale=0.50]{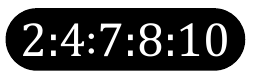}}}
\newcommand{\arsix}{\raisebox{-0.7mm}{\includegraphics[scale=0.50]{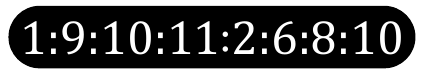}}}

\newcommand{\areight}{\raisebox{-0.7mm}{\includegraphics[scale=0.50]{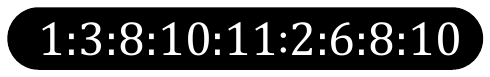}}}
\newcommand{\arnine}{\raisebox{-0.7mm}{\includegraphics[scale=0.50]{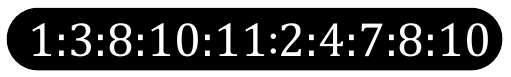}}}

\usepackage{hyperref}

\usepackage{multirow}
\usepackage{pifont}
\usepackage{booktabs}
\usepackage{cleveref}
\crefname{section}{§}{§§}

\usepackage{color}
\usepackage{xcolor}
\usepackage{colortbl}
\usepackage{setspace}
\usepackage{enumitem}
\usepackage{tcolorbox}
\usepackage{hhline}
\usepackage{changepage}
\usepackage[flushleft]{threeparttable}
\usepackage{tabu}
\usepackage{subfig}

\usepackage{blindtext}
\usepackage{tcolorbox}
\usepackage{graphicx}

\newcommand{\xuyuan}[1]{{#1}}
\newcommand{\xingshuo}[1]{{#1}}
\newcommand{\gelei}[1]{{#1}}
\newcommand{\revision}{}
\newcommand{\revisionB}{}
\newcommand{\revisionC}{}
\newcommand{\revisionS}{}

\hypersetup{
  colorlinks   = true,    
  urlcolor     = black,    
  linkcolor    = black,    
  citecolor    = red      
}

\begin{document}

\title{SoK: Rethinking Sensor Spoofing Attacks against Robotic Vehicles \\from a Systematic View}

\author{\rm Yuan Xu\textsuperscript{1}, Xingshuo Han\textsuperscript{1}, Gelei Deng\textsuperscript{1}, Jiwei Li\textsuperscript{2,3}, Yang Liu\textsuperscript{1}, Tianwei Zhang\textsuperscript{1}\\
\textsuperscript{1}Nanyang Technological University, 
\textsuperscript{2}Shannon.AI, 
\textsuperscript{3}Zhejiang University\\
\{xu.yuan, xingshuo.han, gelei.deng, yangliu, tianwei.zhang\}@ntu.edu.sg, jiwei\_li@shannonai.com}

\maketitle

\thispagestyle{plain}
\pagestyle{plain}

\begin{abstract}
Robotic Vehicles (RVs) have gained great popularity over the past few years. Meanwhile, they are also demonstrated to be vulnerable to sensor spoofing attacks. Although a wealth of research works have presented various attacks, some key questions remain unanswered: are these existing works complete enough to cover all the sensor spoofing threats? If not, how many attacks are not explored, and how difficult is it to realize them?

This paper answers the above questions by comprehensively systematizing the knowledge of sensor spoofing attacks against RVs. Our contributions are threefold. (1) We identify seven common attack paths in an RV system pipeline. We categorize and assess existing spoofing attacks from the perspectives of spoofer property, operation, victim characteristic and attack goal. Based on this systematization, we identify 4 interesting insights about spoofing attack designs. (2) We propose a novel action flow model to systematically describe robotic function executions and \revision{unexplored sensor spoofing threats}. With this model, we successfully discover 103 spoofing attack vectors, 26 of which have been verified by prior works, while \xuyuan{77} attacks are never considered. (3) We design two novel attack methodologies to verify the feasibility of newly discovered spoofing attack vectors.
\end{abstract}



\section{Introduction}
\label{sec:intro}
Robotic Vehicles (RVs), such as self-driving cars, automated guided vehicles and drones, enrich our life with myriad scenarios ranging from package delivery, house cleaning to aerial photography. To accomplish these complex missions, an RV system commonly integrates dozens of functions to manage the physical components (i.e., sensors and actuators) and interact with the environments. These functions are constructed as a closed-loop pipeline with various stages \cite{xuyuan-ipdps}: (1) processing sensor data to estimate the states (\textit{perception}); (2) making decisions to achieve the goals (\textit{planning}); (3) taking actions to change the states (\textit{control}). In such cyber-physical systems, sensors are not only the ``eyes'' for RVs to observe and understand the surroundings, but also critical attack surfaces for an external adversary to tamper with the systems and cause catastrophic consequences \cite{sensor-tesla-news,sensor-uber-news,sensor-airplane-news1,sensor-airplane-news2}.

The key to this security threat is \emph{sensor spoofing}, a type of practical physical attack that tricks a victim RV into taking dangerous actions. The adversary first injects fake data into the sensors. Then these malicious data are forwarded to the corresponding perception functions, causing them to generate incorrect state estimates. Such wrong perception results can further affect the subsequent executions in the planning and control stages, and finally lead to undesired hazards. For example, a LiDAR spoofer can craft fake laser points to fool the \textit{object detection} function of an autonomous vehicle, which makes it recognize a non-existent obstacle ahead. This can force the vehicle to brake hard on the highway, causing rear-end collisions and endangering the safety of passengers \cite{spoof-lidar-ccs19}. 
A large number of works have proposed different types of sensor spoofing attacks \cite{sok-gps-ion14,sok-gps-ieee16,sok-gps-cs16,sok-gps-access20,sok-gps-pcs21,sok-camera-access18,sok-camera-arxiv18,sok-camera-cs18,sok-camera-csr20,sok-analog-sc15,sok-analog-sp21}.
We ask the following question: \textit{are existing works complete and in-depth enough for us to understand the fundamentals of sensor spoofing threats, and identify all potential attacks?}

Unfortunately, the answer to the above question is no. Existing works on sensor spoofing attacks fall into two main categories. (1) \emph{Perception-level attacks} \cite{spoof-gps-sensor-ccs12,spoof-gps-sensor-ccs11,spoof-gps-sensor-ion08,spoof-camera-stick-iclr17,anlysis-imu-imece07,anlysis-imu-isie07,anlysis-imu-tie11,spoof-voice-inaudible-ccs17,spoof-voice-inaudible-mobisys17,spoof-voice-inaudible-poster-ccs17,spoof-voice-inaudible-nsdi18,spoof-voice-inaudible-ndss20,spoof-voice-audio-us16,spoof-voice-audio-us18,spoof-voice-audio-sp18,spoof-voice-audio-ndss19,spoof-voice-audio-us20,spoof-voice-audio-icml19,spoof-radar-phdthesis,spoof-camera-stick-arxiv21} target one particular perception function to make it estimate incorrect states. (2) \emph{Vehicle-level attacks} \cite{spoof-gps-drone-vlsid18,spoof-gps-drone-netw19-1,spoof-gps-drone-netw19-2,spoof-gps-uav-us22,spoof-gps-car-us18,spoof-gps-car-us20,spoof-gps-car-iccv21,spoof-gps-car-autosec21,spoof-lidar-bhe15,spoof-lidar-shin,jin2022pla,spoof-lidar-ccs19,spoof-lidar-us20,spoof-lidar-camera-sp21,spoof-lidar-ccs21,spoof-camera-stick-cvpr18,spoof-camera-stick-woot18,spoof-camera-stick-ccs19,kong2020physgan,wang2021dual,cheng2022physical,huang2020universal,wu2020making,xu2020adversarial,spoof-camera-light-ccs21,spoof-camera-raid,yan2022rolling,spoof-camera-light-ccs20,spoof-camera-light-cvpr21,spoof-camera-light-us21,spoof-camera-light-iacr20,spoof-camera-uav-us22,spoof-camera-stick-us21,spoof-camera-stick-us21-cq,spoof-camera-stick-iclr20,spoof-camera-light-woot16,spoof-imu-us15,spoof-imu-eurosp17,spoof-imu-blackhat17,spoof-imu-asiaccs18,spoof-imu-us18,spoof-imu-sp21,spoof-voice-wc19,spoof-voice-inaudible-tdsc21,spoof-ultrosonic-iotj18,spoof-ultrosonic-defcon16,spoof-ultrosonic-ashes19,spoof-radar-tifs21} consider not only fooling the perception function, but also propagating the wrong state estimates towards the subsequent stages and final actions. Both categories of works are only limited to a few specific functions and control flows in an RV system, leaving a large number of unexplored \revision{threats}. This indicates the existence of a knowledge gap about sensor spoofing attacks, and modern RV systems are vulnerable to unknown attacks.     

To bridge this gap, this paper presents a systematic study about sensor spoofing attacks against RV systems. We make three contributions. First, we perform a thorough systematization of sensor spoofing attacks (\cref{sec:review}). Particularly, we identify 7 common attack paths in the RV system pipeline (\cref{sec:2}). Then we categorize existing attacks from 71 relevant papers, covering 6 types of mainstream sensors and 3 types of RV systems. For each attack, we analyze its practicality, aggressivity and stealthiness from four perspectives: spoofer property, spoofing operation, victim characteristic and attack goal. Based on the systematization, we also identify 4 interesting insights that inspire researchers to explore more sophisticated attacks. 

Second, we build a unified action flow model to describe the sensor spoofing attacks, and discover \xuyuan{77} new \revision{unexplored threats} (\cref{sec:3}). 
The key insight of our model is to abstract the spoofing attacks based on their \textit{action flows}, which are defined as end-to-end paths from the sensor data to the RV's final actions. Each robotic function in an action flow can be the attack target, and compromising it could directly or indirectly affect the RV's behaviors. Based on this unified model, we identify 44 action flows and 103 spoofing attack vectors. By analyzing all these 103 attack vectors, we find 26 of them have been realized previously, and they cover all the existing works. 
More importantly, \revisionC{\xuyuan{77} attacks have the potential to cause fatal accidents.} They are never considered in prior works. 

Third, we propose two new attack methodologies to validate the feasibility of the discovered \revision{unexplored threats} (\cref{sec:4}). Specifically, we perform an investigation towards these \xuyuan{77} attack vectors. We find most of them can be easily realized using similar techniques from prior works, while the rest can be categorized into two scenarios: obstacle position altering using the LiDAR spoofer and location altering using the camera/LiDAR spoofer. We design novel approaches to achieve these scenarios. We implement prototypes on the KITTI~\cite{kitti} dataset with the PointRCNN model \cite{shi2019pointrcnn} and ORB-SLAM2 \cite{orb-slam2} simulator to prove these attacks are practical. 

\section{RV System and Spoofing Attack Paths} 
\label{sec:2}

\subsection{RV System Pipeline}
\label{sec:2-attacker-goal}
\begin{figure}[tb]
\centering
\includegraphics[width=1.0\linewidth]{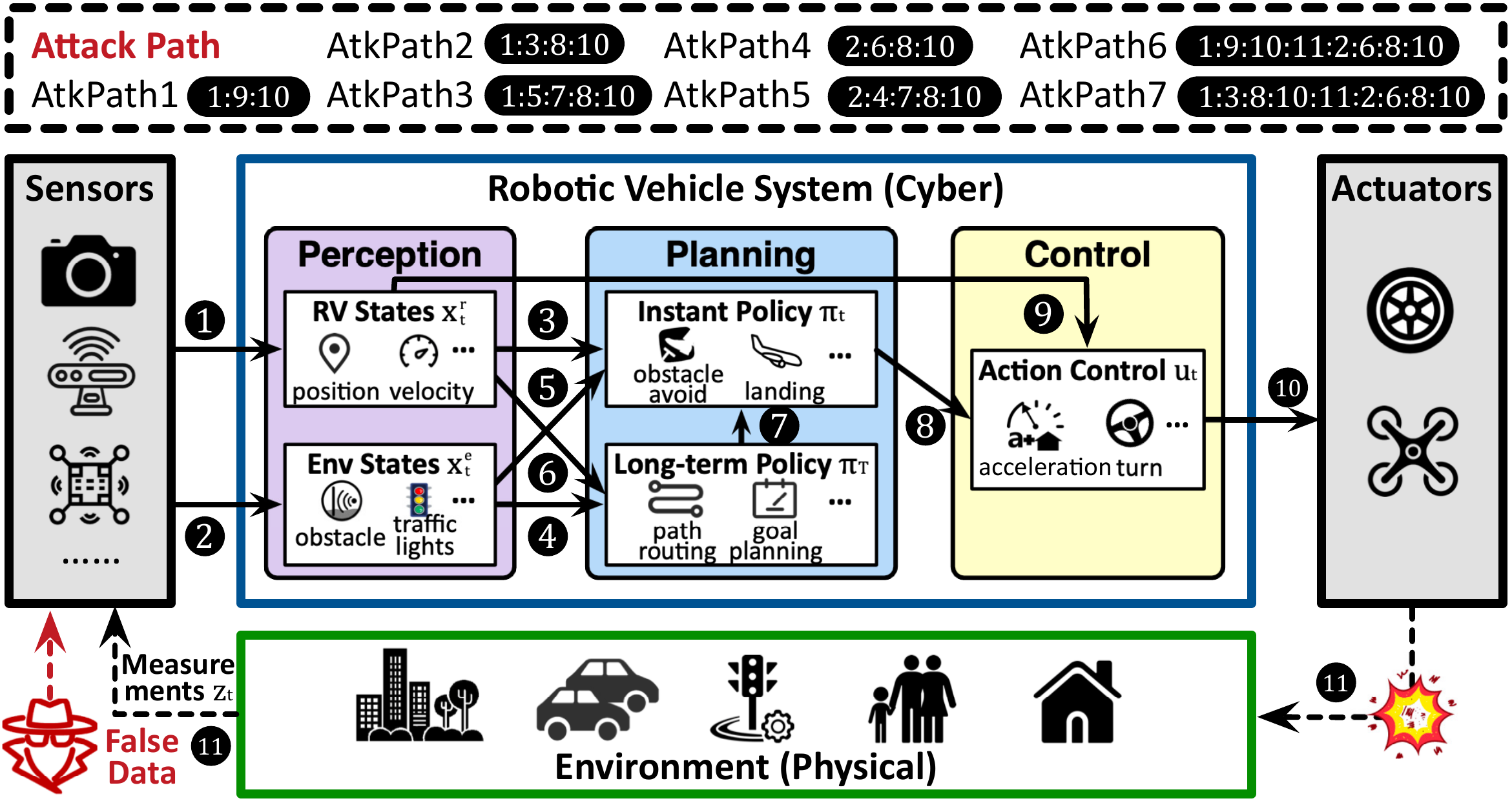}
\vspace{-15pt}
\caption{Overview of an RV system pipeline.}
\vspace{-20pt}
\label{fig:2-atk-model}
\end{figure}

An RV system can be generally modeled as a set of sensor inputs, system states, and control outputs. These three components change over time when the RV takes actions to interact with the environment. They are denoted as $z_t$, $x_t$ and $u_t$ at time $t$, respectively. Figure \ref{fig:2-atk-model} shows the workflow of an RV system. The pipeline is composed of three stages. 

\vspace{3pt}
\noindent\textbf{Perception.} 
The RV estimates its system states $x_t$ from the raw sensor data $z_t$. 
The system states $x_t$ include two parts: $x_t^{r}$ represents RV's operation states (e.g., velocity, position) while $x_t^{e}$ represents the perceived environment states (e.g., nearby obstacles, traffic lanes, pedestrians). 

\vspace{3pt}
\noindent\textbf{Planning.}
To accomplish the mission, the RV needs to make a long-term policy ($\pi_T$) to transit its initial states $x_1$ to the final states $x_T$. During the execution, it keeps computing the instant control policy ($\pi_t$) from $x_t$ to $x_{t+1}$. 
For instance, before a journal, a vehicle first needs to plan a path ($\pi_T$) from the current position ($x_1$) to the destination ($x_T$). When driving along this path, the vehicle needs to ensure that it does not encounter any obstacles and complies with the corresponding traffic laws ($\pi_t$).

\vspace{3pt}
\noindent\textbf{Control.} 
This stage instructs the actuators and drives the RV to interact with the environment. The control outputs $u_t$ depend on the instant policy $\pi_t$ and system states $x_t$. For example, an RV needs to stop or make a turn to avoid the obstacle ahead.

\subsection{Attack Paths}
\label{sec:2-AP}

\bheading{Threat model.}
Following the standard threat model in sensor spoofing works, we assume that the adversary cannot directly access the victim RV, altering its configuration settings or installing malware apps. He can only change the external environment or interfere with the sensor data. The adversary's goal is to tamper with the sensor inputs ($z_t$), which then compromise the \revision{system states} $x_t$ and \revision{control outputs} ($u_t$). 

We analyze the possible attack paths \revision{(AtkPaths)} in the RV pipeline. An attack path describes how fake sensor data can affect the subsequent function executions and results (Figure \ref{fig:2-atk-model}). We first identify five attack paths to compromise the RV:

\begin{itemize}[leftmargin=*]

\item \textbf{\revision{AtkPath1}} (\arone): The spoofer alters the \revision{operation states} $x_t^{r}$ (\one) to destabilize the control stage (\nine), and further cause abnormal  actions $u_t$ (\ten). For example, fake \revisionC{Inertial Measurement Unit (IMU)} data can destabilize a drone and drag it down \cite{spoof-imu-us15}.

\item \textbf{\revision{AtkPath2}} (\artwo): The adversary falsifies the \revision{operation states} $x_t^{r}$ (\one) to influence the \revision{instant policy} $\pi_t$ (\three). A misguided decision-making process will generate \revision{wrong actions} $u_t$ (\eight) and put the RV in danger (\ten). For instance, a drone is forced to land when it is misdirected to a no-fly zone \cite{spoof-gps-drone-vlsid18}.

\item \textbf{\revision{AtkPath3}} (\arthree): The spoofer changes the \revision{operation states} $x_t^{r}$ (\one) to mislead the long-term policy $\pi_T$ (\five) and then the \revision{instant policy} $\pi_t$ (\seven). This continuously controls \revision{RV's actions} $u_t$ (\eight) until it reaches a \revision{malicious final state} $x_T$ (\ten). For example, an adversary can guide the vehicle to a wrong destination by continuously and slightly shifting the GPS location \cite{spoof-gps-car-us18}. 

\item \textbf{\revision{AtkPath4}} (\arfour): This attack path is similar as \revision{AtkPath2}, except that it tampers with the \revision{environment states} $x_t^{e}$. For example, mis-estimating a non-existent obstacle can force the RV to brake hard on the highway \cite{spoof-lidar-ccs19}.

\item \textbf{\revision{AtkPath5}} (\arfive): This attack compromises the \revision{environment states} $x_t^{e}$ to achieve similar consequences as \revision{AtkPath3}. For example, an adversary can spoof the microphone to assign a wrong navigation mission to a vehicle and force it to reach \mbox{a designated destination \cite{spoof-voice-wc19}.}

\end{itemize}

\begin{figure*}[tb]
\centering
\includegraphics[width=0.95\linewidth]{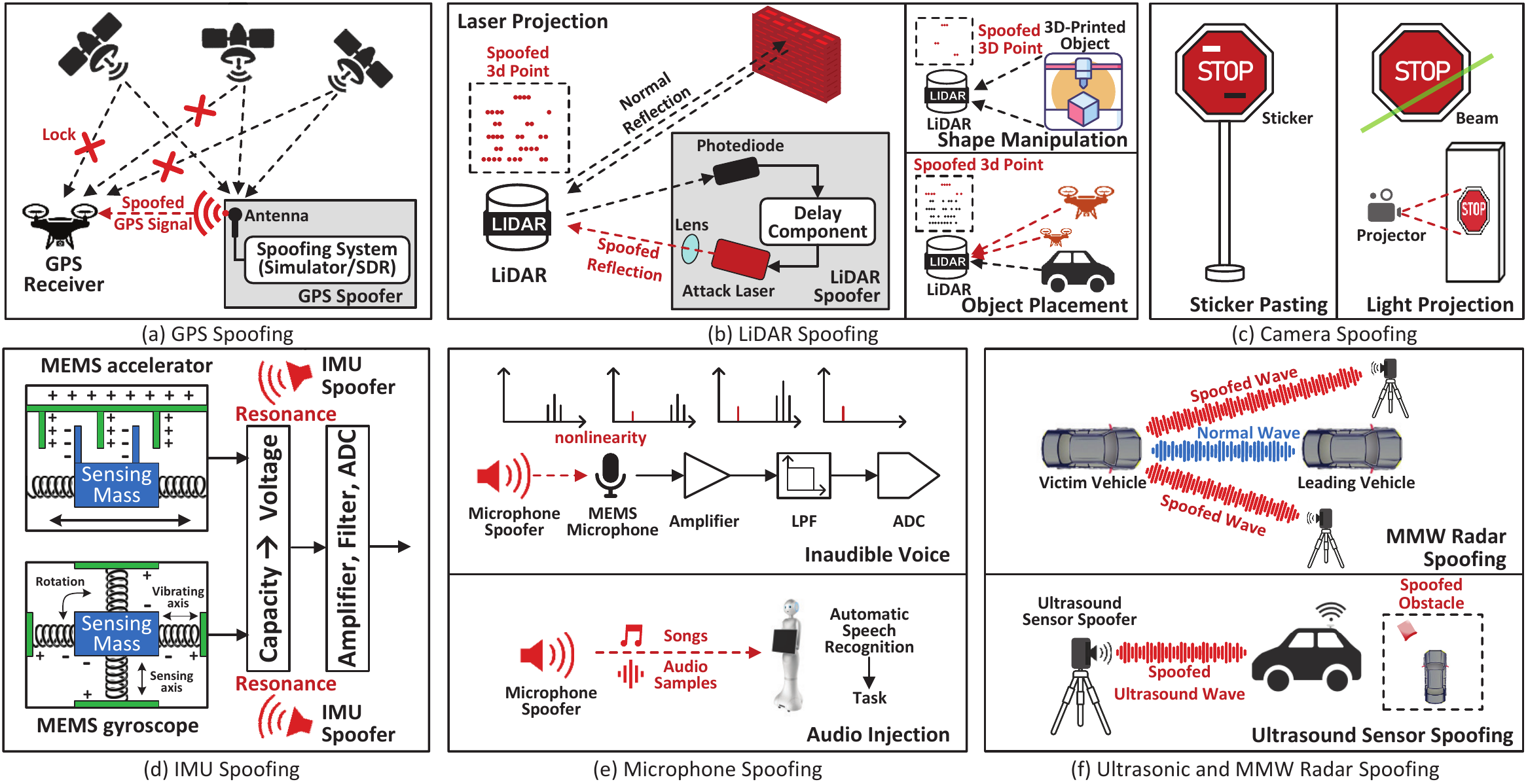}
\vspace{-5pt}
\caption{Illustration of sensor spoofing attacks on six mainstream sensors.}
\vspace{-18pt}
\label{fig:2-spoof-attacks}
\end{figure*}

\xuyuan{
The \revision{environment states} $x_t^{e}$ sensed by the RV changes along with \revision{the operation states} $x_t^{r}$. Once the actions are compromised, the RV perceives the surroundings in unexpected manners and make false decisions. This is reflected in Figure~\ref{fig:2-atk-model} where an adversary can leverage \revision{AtkPath1-3} to achieve the attack result of \revision{AtkPath4-5}. We identify such two attack paths, and call them \mbox{\emph{multi-round attack path}.}
}

\begin{itemize}[leftmargin=*]

\item \textbf{\revision{AtkPath6}} (\arsix): It first performs \revision{AtkPath1} to influence the perceived environment and \revision{sensor measurements} $z_t$ (\one\nine\ten\eleven), and then causes the launch of \revision{AtkPath4} (\two\six\eight\ten). \revisionC{For example, an adversary can control the IMU data to trigger unnecessary motion compensation and generate a blurred image. The blurred images can then induce object misclassification to make the RV take dangerous decisions \cite{spoof-imu-sp21}.}

\item \textbf{\revision{AtkPath7}} (\areight): The adversary falsifies \revision{a malicious position} $x_t^{r}$ and causes the change of \revision{sensor measurements} $z_t$ (\one\three\eight\ten\eleven), which then leads to the occurrence of \revision{AtkPath4}  (\two\six\eight\ten). 
\revisionC{For example, when multiple traffic lights exist in the camera view, a counterfeit position can make the vehicle confused about different traffic lights in the region-of-interests (ROI) and take unexpected actions because ROI utilizes position to narrow the detection scope in the sensor \cite{spoof-gps-car-autosec21}.}

\end{itemize}

\xuyuan{
Note that there can be some other two-round attack paths from Figure \ref{fig:2-atk-model}, including \revision{AtkPath3+4} and \revision{AtkPath1/2/3+5}. We observe that \revision{AtkPath1}, \revision{AtkPath3} and \revision{AtkPath4} aim to trigger \emph{instant} abnormal events and cause malicious actions, while \revision{AtkPath2} and \revision{AtkPath5} focus on continuously spoofing sensors to change the \emph{long-term} goals. Thus, it is hard to combine them in practice. \revision{AtkPath2+5} (i.e. \arnine) can be attacked in theory if the vehicle performs different missions according to its position. However, we cannot find any works about these attacks, and will not discuss them in this paper.
}


\subsection{Systematization Scope}
This paper focuses on sensor spoofing attacks against RVs. Although there are almost 400 types of sensors on record \cite{sensor-list}, we consider six mainstream sensors: GPS, LiDAR, camera, IMU, microphone, and ultrasonic sensor/MMW radar. They are fully or partially integrated into modern RVs to assist them for planning and control. Spoofing attacks on other sensors, e.g. temperature sensor \cite{spoof-temperature-ccs19}, pressure sensor \cite{spoof-pressure-us10,spoof-pressure-asiaccs21} and magnetic sensor \cite{spoof-magnetic-ccs15,spoof-magnetic-ches13}, are beyond the scope of this paper. Attacks against other cyber-physical systems (e.g., smart-home speakers \cite{abdullah2021sok}, medical devices \cite{park2016ain}) are not studied either. 

We target three types of popular RVs (autonomous vehicles, drones, and automated guided vehicles), which are widely adopted in our daily life. These RV systems follow the standard modular design and pipeline described in \cref{sec:2-attacker-goal}. We do not consider the end-to-end robotic systems which utilize single machine learning models to directly output the control command from the sensor data \cite{cheng2018end,tampuu2020survey}, as they are generally explored in academia, and not ready for commercial use. Besides, we mainly focus on the vehicle-level attacks \cite{spoof-gps-drone-vlsid18,spoof-gps-drone-netw19-1,spoof-gps-drone-netw19-2,spoof-gps-uav-us22,spoof-gps-car-us18,spoof-gps-car-us20,spoof-gps-car-iccv21,spoof-gps-car-autosec21,spoof-lidar-bhe15,spoof-lidar-shin,jin2022pla,spoof-lidar-ccs19,spoof-lidar-us20,spoof-lidar-camera-sp21,spoof-lidar-ccs21,spoof-camera-stick-cvpr18,spoof-camera-stick-woot18,spoof-camera-stick-ccs19,kong2020physgan,wang2021dual,cheng2022physical,huang2020universal,wu2020making,xu2020adversarial,spoof-camera-raid,yan2022rolling,spoof-camera-light-ccs20,spoof-camera-light-cvpr21,spoof-camera-light-ccs21,spoof-camera-light-us21,spoof-camera-light-iacr20,spoof-camera-uav-us22,spoof-camera-stick-us21,spoof-camera-stick-us21-cq,spoof-camera-stick-iclr20,spoof-camera-light-woot16,spoof-imu-us15,spoof-imu-eurosp17,spoof-imu-blackhat17,spoof-imu-asiaccs18,spoof-imu-us18,spoof-imu-sp21,spoof-voice-wc19,spoof-voice-inaudible-tdsc21,spoof-ultrosonic-iotj18,spoof-ultrosonic-defcon16,spoof-ultrosonic-ashes19,spoof-radar-tifs21} \xuyuan{, i.e., an end-to-end attack causing malicious actions from the spoofed perceived states to the planning and control subsystems}. Perception-level attacks \cite{spoof-gps-sensor-ccs12,spoof-gps-sensor-ccs11,spoof-gps-sensor-ion08,spoof-camera-stick-iclr17,anlysis-imu-imece07,anlysis-imu-isie07,anlysis-imu-tie11,spoof-voice-inaudible-ccs17,spoof-voice-inaudible-mobisys17,spoof-voice-inaudible-poster-ccs17,spoof-voice-inaudible-nsdi18,spoof-voice-inaudible-ndss20,spoof-voice-audio-us16,spoof-voice-audio-us18,spoof-voice-audio-sp18,spoof-voice-audio-ndss19,spoof-voice-audio-us20,spoof-voice-audio-icml19,spoof-radar-phdthesis} can be regarded as the preliminary step for vehicle-level attacks, and will not be particularly discussed in our systematization.

\gelei{While jamming attacks~\cite{jamming-gps-jss19,jamming-gps-sensor21} can cause malfunctions of RVs, we exclude them from this work due to two main considerations. First, sensor jamming attacks crudely block the perceived data instead of arbitrarily modifying the sensor readings. This gives the adversary less flexibility to mislead the RVs in a more precise way. Second, jamming attacks are generally not stealthy and \revisionC{easy to detect}~\cite{sun21platoonjamming}. Existing works have demonstrated solutions to defeat jamming attacks on various sensors, such as Radar~\cite{orlando2016novel, greco2008radar}, GPS~\cite{axell2015jamming, moussa2017enhanced}, and ultrasonic sensor~\cite{lee2019securing}. Meanwhile, we do not consider cyber attacks against RVs (e.g., software and ROS vulnerabilities \cite{xuyuan-raid,SP-robot-controller} and in-vehicle networks \cite{9519391,bloom2021weepingcan,bhatia2021evading}, DNN backdoor \cite{han2022clean}, communication protocols \cite{gelei-raid,kim2021pgfuzz}, and side-channel leakage \cite{luo2020stealthy}) in this work because they do not directly target on-vehicle sensors.}


\noindent\textbf{Comparisons with existing surveys.}
A few works also conduct surveys related to RV or sensor security. However, they are significantly different from this paper. (1) \emph{Scope}: some papers focus on the general security and safety problems in specific RV systems, e.g., drones \cite{sok-drone-jds15,spoof-gps-drone-defcon16,sok-drone-chi17,sok-drone-tcps17,sok-drone-ioa17,sok-drone-sp21,sok-drone-mna20,sok-drone-icca20,sok-drone-secon20}, autonomous vehicles \cite{spoof-lidar-bhe15,sok-car-tits17,sok-car-isis18,sok-car-irc19,sok-car-ieee20,sok-car-access20,qayyum2020securing,kim2021cybersecurity,sok-car-iotj21,sok-car-tii21,sok-car-arxiv22}. Some papers just target one type of sensor spoofing, e.g., GPS \cite{sok-gps-ion14,sok-gps-ieee16,sok-gps-cs16,sok-gps-access20,sok-gps-pcs21}, camera \cite{sok-camera-access18,sok-camera-arxiv18,sok-camera-cs18,sok-camera-csr20} or microphone \cite{sok-analog-sc15,abdullah2021sok}). 
Differently, we explore various types of sensor spoofing attacks against different RV systems. (2) \emph{Contribution}: we provide a deeper analysis and categorization on spoofing attacks in a systematic way. We identify the common attack paths from the RV system pipeline and assess existing works from different perspectives. We build an action flow model, which can not only cover existing attacks but also disclose new \revision{unexplored threats}. We also design and prototype new spoofing attack approaches. These are rarely performed in prior works. Yan et al. \cite{sok-analog-sp21} also introduced a sensor security model to describe spoofing attacks and predict new vectors. It focuses on the underlying signal processing mechanisms in each sensor at the perception level. On the contrary, our action flow model is based on robotic function executions and interactions at the vehicle level, which is complementary to \cite{sok-analog-sp21}.

\begin{table*}[tb]
\vspace{-15pt}
\caption{Systematization of sensor spoofing attacks.}
\vspace{-6pt}
\centering
\resizebox{1.00\linewidth}{!}{
\begin{tabular}{|cc|cccc|cc|cc|cc|c|}
\rowcolor{gray!40}
\hline
\multicolumn{2}{|c|}{\multirow{2}{*}{\cellcolor{gray!40}}} & \multicolumn{4}{c|}{\textbf{Spoofer Property}} & \multicolumn{2}{c|}{\textbf{Spoofing Operation}} & \multicolumn{2}{c|}{\textbf{Victim Characteristic}} & \multicolumn{2}{c|}{\textbf{Attack Goal}} & {\cellcolor{gray!40}} \\ 
\hhline{|>{\arrayrulecolor{gray!40}}-->{\arrayrulecolor{black}}---------->{\arrayrulecolor{gray!40}}->{\arrayrulecolor{black}}|}
\multicolumn{2}{|c|}{\multirow{-2}{*}{\cellcolor{gray!40}\textbf{Spoofing Techniques}}} & \multicolumn{1}{c|}{\cellcolor{gray!40}Cost} & \multicolumn{1}{c|}{\cellcolor{gray!40}Size} & \multicolumn{1}{c|}{\cellcolor{gray!40}Signal} & \multicolumn{1}{c|}{\cellcolor{gray!40}Recog.} & \multicolumn{1}{c|}{\cellcolor{gray!40}Range} & \multicolumn{1}{c|}{\cellcolor{gray!40}Exposure} & \multicolumn{1}{c|}{\cellcolor{gray!40}\textcolor{gray!40}{xx}Type\textcolor{gray!40}{xx}} & \multicolumn{1}{c|}{\cellcolor{gray!40}Scenario} & \multicolumn{1}{c|}{\cellcolor{gray!40}State} & \multicolumn{1}{c|}{\cellcolor{gray!40}Attack Path}& \multirow{-2}{*}{\cellcolor{gray!40}\textbf{Paper}} \\
\hline
\multicolumn{2}{|c|}{\multirow{4}{*}{GPS Spoofing}}	& 
	\multicolumn{1}{c|}{\multirow{4}{*}{\$\$/\$\$\$}} & \multicolumn{1}{c|}{\multirow{4}{*}{S1/S2}} & \multicolumn{1}{c|}{\multirow{4}{*}{\gps}} & \multicolumn{1}{c|}{\multirow{4}{*}{False}} & \multicolumn{1}{c|}{\multirow{4}{*}{R2}} & \multicolumn{1}{c|}{\multirow{4}{*}{Active}} & \multicolumn{1}{c|}{\multirow{2}{*}{\drone \car}} & \multirow{2}{*}{\outdoor} & \multicolumn{1}{c|}{\multirow{2}{*}{Position}} & \revision{AtkPath2} & \cite{spoof-gps-drone-vlsid18,spoof-gps-drone-netw19-1,spoof-gps-drone-netw19-2,spoof-gps-uav-us22} \\
	\cline{12-13}
	\multicolumn{1}{|c}{} & \multicolumn{1}{c|}{} & \multicolumn{1}{c|}{} & \multicolumn{1}{c|}{} & \multicolumn{1}{c|}{} & \multicolumn{1}{c|}{} & \multicolumn{1}{c|}{} & \multicolumn{1}{c|}{} & \multicolumn{1}{c|}{} & \multicolumn{1}{c|}{} & \multicolumn{1}{c|}{} & \revision{AtkPath3} & \cite{spoof-gps-uav-us22,spoof-gps-car-us18,spoof-gps-car-us20} \\
\cline{9-13}
	\multicolumn{1}{|c}{} & \multicolumn{1}{c|}{} & \multicolumn{1}{c|}{} & \multicolumn{1}{c|}{} & \multicolumn{1}{c|}{} & \multicolumn{1}{c|}{} & \multicolumn{1}{c|}{} & \multicolumn{1}{c|}{} & \multicolumn{1}{c|}{\multirow{2}{*}{\car}} & \multirow{2}{*}{\outdoor} & \multicolumn{1}{c|}{\multirow{2}{*}{Object}} & \revision{AtkPath6} & \cite{spoof-gps-car-iccv21} \\
	\cline{12-13}
	\multicolumn{1}{|c}{} & \multicolumn{1}{c|}{} & \multicolumn{1}{c|}{} & \multicolumn{1}{c|}{} & \multicolumn{1}{c|}{} & \multicolumn{1}{c|}{} & \multicolumn{1}{c|}{} & \multicolumn{1}{c|}{} & \multicolumn{1}{c|}{} & \multicolumn{1}{c|}{} & \multicolumn{1}{c|}{} & \revision{AtkPath7} & \cite{spoof-gps-car-autosec21} \\
\hline
\multicolumn{1}{|c|}{\multirow{3}{*}{\begin{tabular}[c]{@{}c@{}}LiDAR\\	Spoofing\end{tabular}}}	& laser projection & 
	\multicolumn{1}{c|}{\$\$\$} & \multicolumn{1}{c|}{S2} & \multicolumn{1}{c|}{\laser} & \multicolumn{1}{c|}{False} & \multicolumn{1}{c|}{R2} & \multicolumn{1}{c|}{Active} & \multicolumn{1}{c|}{\car} & \outdoor & \multicolumn{1}{c|}{Object} & \revision{AtkPath4} & \cite{spoof-lidar-bhe15,spoof-lidar-shin,jin2022pla,spoof-lidar-ccs19,spoof-lidar-us20} \\ 
\cline{2-13} 
\multicolumn{1}{|c|}{}	& shape manipulation & 
	\multicolumn{1}{c|}{\$\$} & \multicolumn{1}{c|}{S1} & \multicolumn{1}{c|}{\shape} & \multicolumn{1}{c|}{True} & \multicolumn{1}{c|}{R2}	& \multicolumn{1}{c|}{Passive} & \multicolumn{1}{c|}{\car} & \outdoor & \multicolumn{1}{c|}{Object} & \revision{AtkPath4} & \cite{spoof-lidar-camera-sp21}\\
\cline{2-13} 
\multicolumn{1}{|c|}{}	& object placement & 
	\multicolumn{1}{c|}{\$\$} & \multicolumn{1}{c|}{S1} & \multicolumn{1}{c|}{\shape} & \multicolumn{1}{c|}{True} & \multicolumn{1}{c|}{R2}	& \multicolumn{1}{c|}{Passive} & \multicolumn{1}{c|}{\car} & \outdoor & \multicolumn{1}{c|}{Object} & \revision{AtkPath4} & \cite{spoof-lidar-ccs21} \\
\hline
\multicolumn{1}{|c|}{\multirow{9}{*}{\begin{tabular}[c]{@{}c@{}}Camera\\ Spoofing\end{tabular}}} & \multirow{3}{*}{sticker pasting} & 
	\multicolumn{1}{c|}{\multirow{3}{*}{\$}} & \multicolumn{1}{c|}{\multirow{3}{*}{S1}} & \multicolumn{1}{c|}{\multirow{3}{*}{\sticker}} & \multicolumn{1}{c|}{\multirow{3}{*}{True}} & \multicolumn{1}{c|}{\multirow{3}{*}{R2}} & \multicolumn{1}{c|}{\multirow{3}{*}{Passive}}  & \multicolumn{1}{c|}{\car} & \outdoor & \multicolumn{1}{c|}{Object} & \revision{AtkPath4} & \cite{spoof-camera-stick-cvpr18,spoof-camera-stick-woot18,spoof-camera-stick-ccs19,kong2020physgan,wang2021dual,cheng2022physical,huang2020universal,wu2020making,xu2020adversarial} \\
\cline{9-13} 
	\multicolumn{1}{|c|}{} & \multicolumn{1}{c|}{} & \multicolumn{1}{c|}{} & \multicolumn{1}{c|}{} & \multicolumn{1}{c|}{} & \multicolumn{1}{c|}{} & \multicolumn{1}{c|}{} & \multicolumn{1}{c|}{} & \multicolumn{1}{c|}{\car} & \outdoor & \multicolumn{1}{c|}{Object} & \revision{AtkPath5} & \cite{spoof-camera-stick-iclr20} \\
\cline{9-13} 
	\multicolumn{1}{|c|}{} & \multicolumn{1}{c|}{} & \multicolumn{1}{c|}{} & \multicolumn{1}{c|}{} & \multicolumn{1}{c|}{} & \multicolumn{1}{c|}{} & \multicolumn{1}{c|}{} & \multicolumn{1}{c|}{} & \multicolumn{1}{c|}{\car} & \outdoor & \multicolumn{1}{c|}{Lane} & \revision{AtkPath4} & \cite{spoof-camera-stick-us21,spoof-camera-stick-us21-cq} \\
\cline{2-13} 
\multicolumn{1}{|c|}{} & \multicolumn{1}{c|}{\multirow{4}{*}{light projection}} & 
	\multicolumn{1}{c|}{\multirow{4}{*}{\$/\$\$}}& \multicolumn{1}{c|}{\multirow{4}{*}{S1}} & \multicolumn{1}{c|}{\multirow{4}{*}{\laser}} & \multicolumn{1}{c|}{False} & \multicolumn{1}{|c|}{R2} & \multicolumn{1}{c|}{Active} & \multicolumn{1}{c|}{\car} & \outdoor & \multicolumn{1}{c|}{Object} & \revision{AtkPath4} & \cite{spoof-camera-light-ccs21} \\
\cline{6-13}
	\multicolumn{1}{|c|}{\multirow{3}{*}{}} & \multicolumn{1}{c|}{\multirow{3}{*}{}} & \multicolumn{1}{c|}{} & \multicolumn{1}{c|}{} & \multicolumn{1}{c|}{} & \multicolumn{1}{c|}{\multirow{3}{*}{True}} & \multicolumn{1}{c|}{\multirow{3}{*}{R2}} & \multicolumn{1}{c|}{\multirow{3}{*}{Active}} & \multicolumn{1}{c|}{\drone\car} & \outdoor & \multicolumn{1}{c|}{Object} & \revision{AtkPath4} & \cite{spoof-camera-raid,yan2022rolling,spoof-camera-light-ccs20,spoof-camera-light-iacr20,spoof-camera-light-cvpr21,spoof-camera-light-us21,spoof-camera-uav-us22} \\
\cline{9-13}
	\multicolumn{1}{|c|}{} & \multicolumn{1}{c|}{} & \multicolumn{1}{c|}{} & \multicolumn{1}{c|}{} & \multicolumn{1}{c|}{} & \multicolumn{1}{c|}{} & \multicolumn{1}{c|}{} & \multicolumn{1}{c|}{} & \multicolumn{1}{c|}{\car} & \outdoor & \multicolumn{1}{c|}{Lane} & \revision{AtkPath4} & \cite{spoof-camera-light-iacr20,spoof-camera-light-ccs20} \\
\cline{9-13}
	\multicolumn{1}{|c|}{} & \multicolumn{1}{c|}{} & \multicolumn{1}{c|}{} & \multicolumn{1}{c|}{} & \multicolumn{1}{c|}{} & \multicolumn{1}{c|}{} & \multicolumn{1}{c|}{} & \multicolumn{1}{c|}{} & \multicolumn{1}{c|}{\drone} & \outdoor & \multicolumn{1}{c|}{Velocity} & \revision{AtkPath1} & \cite{spoof-camera-light-woot16} \\
\hline
\multicolumn{2}{|c|}{\multirow{2}{*}{IMU Spoofing}}	&
	\multicolumn{1}{c|}{\multirow{2}{*}{\$\$}} & \multicolumn{1}{c|}{\multirow{2}{*}{S2}} & \multicolumn{1}{c|}{\multirow{2}{*}{\voice}} & \multicolumn{1}{c|}{\multirow{2}{*}{True}} & \multicolumn{1}{c|}{\multirow{2}{*}{R2}} & \multicolumn{1}{c|}{\multirow{2}{*}{Active}} & \multicolumn{1}{c|}{\mb \drone} & \indoor\,\outdoor & \multicolumn{1}{c|}{Velocity} & \revision{AtkPath1} & \cite{spoof-imu-us15,spoof-imu-eurosp17,spoof-imu-blackhat17,spoof-imu-asiaccs18,spoof-imu-us18} \\
\cline{9-13}
	\multicolumn{1}{|c}{} & \multicolumn{1}{c|}{} & \multicolumn{1}{c|}{} & \multicolumn{1}{c|}{} & \multicolumn{1}{c|}{} & \multicolumn{1}{c|}{} & \multicolumn{1}{c|}{} & \multicolumn{1}{c|}{} & \multicolumn{1}{c|}{\car} & \outdoor & \multicolumn{1}{c|}{Object} & \revision{AtkPath6} & \cite{spoof-imu-sp21} \\
\hline
\multicolumn{1}{|c|}{\multirow{2}{*}{\begin{tabular}[c]{@{}c@{}}Microphone \\ Spoofing\end{tabular}}} & inaudible voice & 
	\multicolumn{1}{c|}{\$\$\$} & \multicolumn{1}{c|}{S2} & \multicolumn{1}{c|}{\voice} & \multicolumn{1}{c|}{False} & \multicolumn{1}{c|}{R2}	& \multicolumn{1}{c|}{Active} & \multicolumn{1}{c|}{\car} & \outdoor & \multicolumn{1}{c|}{Mission Goal} & \revision{AtkPath5} & \cite{spoof-voice-wc19,spoof-voice-inaudible-tdsc21} \\
\cline{2-13} 
\multicolumn{1}{|c|}{}	& audio injection & 
	\multicolumn{1}{c|}{\$} & \multicolumn{1}{c|}{S1} & \multicolumn{1}{c|}{\voice} & \multicolumn{1}{c|}{False} & \multicolumn{1}{c|}{R2} & \multicolumn{1}{c|}{Passive} & \multicolumn{1}{c|}{\car} & \outdoor & \multicolumn{1}{c|}{Mission Goal} & \revision{AtkPath5} & \cite{spoof-voice-wc19}\\
\hline
\multicolumn{2}{|c|}{Ultrasonic Sensor Spoofing} & 
	\multicolumn{1}{c|}{\$} & \multicolumn{1}{c|}{S1} & \multicolumn{1}{c|}{\voice} & \multicolumn{1}{c|}{False} & \multicolumn{1}{c|}{R1} & \multicolumn{1}{c|}{Active} & \multicolumn{1}{c|}{\car} & \indoor\,\outdoor & \multicolumn{1}{c|}{Object} & \revision{AtkPath4} & \cite{spoof-ultrosonic-iotj18,spoof-ultrosonic-defcon16} \\ 
\hline
\multicolumn{2}{|c|}{\multirow{2}{*}{MMW Radar Spoofing}} & 
	\multicolumn{1}{c|}{\multirow{2}{*}{\$\$\$}} & \multicolumn{1}{c|}{\multirow{2}{*}{S2}} & \multicolumn{1}{c|}{\multirow{2}{*}{\wave}} & \multicolumn{1}{c|}{\multirow{2}{*}{False}} & \multicolumn{1}{c|}{\multirow{2}{*}{R2}} & \multicolumn{1}{c|}{\multirow{2}{*}{Active}} & \multicolumn{1}{c|}{\multirow{2}{*}{\car}} & \multirow{2}{*}{\outdoor} & \multicolumn{1}{c|}{\multirow{2}{*}{Object}} & \revision{AtkPath4} & \cite{spoof-ultrosonic-defcon16,spoof-ultrosonic-ashes19,spoof-radar-tifs21}\\ 
\cline{12-13} 
	\multicolumn{1}{|c}{} & \multicolumn{1}{c|}{} & \multicolumn{1}{c|}{} & \multicolumn{1}{c|}{} & \multicolumn{1}{c|}{} & \multicolumn{1}{c|}{} & \multicolumn{1}{c|}{} & \multicolumn{1}{c|}{} & \multicolumn{1}{c|}{} & \multicolumn{1}{c|}{} & \multicolumn{1}{c|}{} & \revision{AtkPath5} & \cite{spoof-radar-tifs21} \\
\hline
\end{tabular}}
\begin{adjustwidth}{0.5cm}{0.5cm}
\begin{tablenotes}[para,flushleft]
\footnotesize
	\centering \gps\;Satellite Signal \quad \laser\;Visible light or infrared \quad \sticker\;2D sticker \quad \shape\;3D object \quad \voice\;Audible sound or ultrasound \quad \wave\;RF waves
	\centering \car\;Autonomous Driving \quad \drone\;Drone \quad \mb\;Mobile Base \quad \textbf{Recog.}\;Recognizability \quad \indoor\;Indoor \quad \outdoor\;OutdoorInjection
\end{tablenotes}
\end{adjustwidth}
\label{table:spoofing-compare}
\vspace{-15pt}
\end{table*}

\section{Systematization of Existing Attacks}
\label{sec:review}
We first present our categorization methodology (\cref{sec:3-1}). Then we perform \revisionC{a} literature review about sensor spoofing attacks (\cref{sec:3-gps}-\cref{sec:3-radar}). Finally, we draw some interesting insights from the systematization (\cref{sec:2-threat-model}). Table \ref{table:spoofing-compare} lists the summary of these works, and Figure \ref{fig:2-spoof-attacks} illustrates the basic mechanism of each attack. 

\subsection{Systematization Methodology}
\label{sec:3-1}


\noindent\textbf{\revision{1) Spoofer Property.}} 
An adversary needs a spoofer to interfere with the victim RV. The spoofers in different attacks can have distinct properties, which determine the attack cost, feasibility and stealthiness. We evaluate four properties. \textbf{a) Cost:} this is the price to purchase or set up the spoofer.
We consider three levels: less than \$100 (\$); between \$100 and \$1000 (\$\$);  more than \$1000 (\$\$\$).  \textbf{b) Size:} this denotes the physical size of the spoofer. It is easier and stealthier to perform attacks with small-size spoofers. We consider two types: non-portable -- larger than a mug (S2); portable -- smaller than a mug (S1). \textbf{c) Signal type:} 
satellite signal (\gps\,); visible or infrared light (\laser\,); 2D sticker (\sticker\,); 3D object (\shape\,); audible sound or ultrasound (\voice\,); and RF waves (\wave\,). \textbf{d) Recognizability:} this denotes whether the spoofer or spoofed signal can be noticed by the victim user (True), or can conceal themselves in the environment (False).

\noindent\textbf{\revision{2) Spoofing Operation.}} 
We consider different operations the adversary performs to attack the RV. \textbf{a) Range}: 
the minimal distance between the spoofer and RV required for effective interference -  
larger than 5m (R2); smaller than 5m (R1). \textbf{b) Exposure:} this indicates whether the adversary needs to actively expose himself to perform attacks (A) or passively mislead the sensors (P). 

\noindent\textbf{\revision{3) Victim Characteristic.}} 
We also assess the attacks based on two characteristics of the victim RVs. 
\textbf{a) Type}: 
autonomous vehicle (\car\,) \revision{with various sensors and strict compliance to traffic rules}; drone (\drone\,) \revision{utilizing IMU and quadrotors for stability}; automated guided vehicle (\mb\,) \revision{with limited low-end sensors due to cost constraints}. 
\textbf{b) Scenario}: 
the indoor scenario (\indoor\,) or outdoor scenario (\outdoor\,). 
\revisionC{There are significant differences between indoor and outdoor RV, which must be taken into account when designing security systems for these vehicles. In indoor environments, GPS localization is not feasible, the RV typically relies on Simultaneous Localization And Mapping (SLAM) or IMUs for localization. Additionally, indoor RVs also tend to run slower than their outdoor counterparts. This can actually be advantageous in terms of stability and \mbox{result in less crash damage in certain attack scenarios.}}

\noindent\textbf{\revision{4) Attack Goal.}} 
We consider the attack goal from two dimensions. 
\textbf{a) Compromised state:} various states can be attacked, including position, nearby obstacles (Object), the traffic lanes (Lane), velocity and mission goals. 
\textbf{b) Attack path}: we identify the attack path (\revision{AtkPath1-7}) exploited by the adversary, as discussed in \cref{sec:2-AP}

\subsection{GPS Spoofing Attacks}
\label{sec:3-gps}

The Global Position System (GPS) is widely integrated into outdoor RVs for localization. 
The GPS receiver calculates its position based on the information received from multiple satellites, including pseudorange and navigation messages. A navigation message consists of the transmission time of the code epoch and the satellite position at that time. 

The lack of signal authentication makes GPS vulnerable to spoofing attacks \cite{gps-spoof-report}. 
As shown in Figure \ref{fig:2-spoof-attacks}(a), the adversary first uses a GPS spoofer to transmit false GPS signals with strong power to the victim GPS receiver. These fabricated signals lure the victim to lose track of the satellites and lock onto the attacker's signals. Next, the adversary can manipulate the GPS receiver by either adjusting the apparent pseudorange to the satellite \cite{spoof-gps-sensor-ion08} or modifying the navigation messages \cite{spoof-gps-sensor-ccs12,spoof-gps-sensor-ccs11}. The counterfeit signal is then sent to the victim GPS receiver. 

\noindent\textbf{\revision{1) Spoofer Property.}} 
There are two common types of GPS spoofers: GPS simulator and Software Defined Radio (SDR). The GPS simulator is heavy, expensive but more powerful. Generally, a complete digital GPS simulator with the multi-GNSS capability is as big as a computer server (size: S2) with the price between \$20,000 and \$50,000 (cost: \$\$\$) \cite{sok-gps-cs16}. Under different configurations, the simulator can simulate from only 10 satellite signals (signal: \gps) at one time (e.g. WelNavigate GS72 \cite{gps-spoofer-welnavigate}) to 64 simultaneous signals and multiple GNSS systems (e.g. Orolia GSG 5/6 series \cite{gps-spoofer-orolia}). In comparison, the SDR spoofer is more popular due to its low-cost, easy-operation and white-box features. 
An attack is successfully demonstrated against autonomous vehicles in \cite{spoof-gps-car-us18} with a pen-size SDR spoofer at the price of \$223 (cost: \$\$, size: S1). 
The GPS spoofer can be hidden in an adversarial vehicle to emit the imperceptible but effective fake GPS signals \cite{spoof-gps-car-us18}. Thus, it is hard for the victim to notice the attack. (recog.: False).

\noindent\textbf{\revision{2) Spoofing Operation.}} 
Both the GPS simulator and SDR spoofer need to transfer the counterfeit satellite signals to the victim's GPS receiver (exposure: A). 
They can be launched at longer distances (e.g., 40m away \cite{spoof-gps-car-us18}) from the victim RV (range: R2). 

\noindent\textbf{\revision{3) Victim Characteristic.}} 
GPS spoofing has been realized to attack drones \cite{spoof-gps-drone-vlsid18,spoof-gps-drone-netw19-1,spoof-gps-drone-netw19-2,sok-gps-pcs21} and autonomous vehicles \cite{spoof-gps-car-us18,spoof-gps-car-us20} in the outdoor scenario since the satellite signals would be blocked by walls (type: \drone\car, scenario: \outdoor). 

\noindent\textbf{\revision{4) Attack Goal.}} 
The adversary compromises the position of the victim RV (state: Position). This can incur additional effects for the subsequent functions with different attack paths. (1) \revision{AtkPath2}: Some attacks were proposed to control DJI drones to enter or leave a no-fly zone \cite{spoof-gps-drone-vlsid18,spoof-gps-drone-netw19-1}. Once reaching the coordinates of a no-fly zone, the drone has to perform an emergency landing, which is unexpected. A counterfeit position can also cause a DJI drone to fly to an incorrect destination after entering the return-to-home mode \cite{spoof-gps-drone-netw19-2}. (2) \revision{AtkPath3}: Zeng et al. \cite{spoof-gps-car-us18} introduced an attack, which continuously and slightly shifts the GPS position to manipulate the road navigation system of an autonomous vehicle. Then the fake navigation route will match the shape of the actual roads, and induce the vehicle to a dangerous destination. Shen et al. \cite{spoof-gps-car-us20} 
proposed the \emph{off-road} and \emph{wrong-way} attacks, which perform continuous GPS spoofing with large lateral deviations to mislead the autonomous vehicle to drive off the road or onto the opposite lane. 

The adversary can also change the semantic information of the objects based on the spoofed position (state: Object). Li et al. \cite{spoof-gps-car-iccv21} designed an attack against the motion compensation mechanism, which uses the GPS information to fix LiDAR distortions. By falsifying the positions with the GPS spoofer, the adversary can compromise the LiDAR-based object detection function and make safety-critical objects undetectable by the victim vehicle (\revision{AtkPath: 6}). 
Tang et al.~\cite{spoof-gps-car-autosec21} proposed to manipulate the location with GPS spoofing to affect the position of ROI in the traffic light detection function, thus leading the victim vehicle to run \revisionC{a} red light (\revision{AtkPath: 7}).

\xuyuan{
GPS can also be used to estimate the velocity through differentiating two consecutive positions \cite{gps-vel-1,gps-vel-2}. However, most commercial RVs use IMU or camera as the velocity estimator rather than GPS due to the accuracy. Thus, this paper does not consider this opportunity. 
}

\subsection{LiDAR Spoofing Attacks}

A LiDAR sensor is used to measure the distance from the RV to surroundings by the Time-of-Flight (ToF) method, i.e. firing rapid laser pulses and capturing the reflected light using photodiodes. 
With such information, the RV can recognize the shape and position of any object in the form of point clouds. 

\xingshuo{Three basic techniques to conduct LiDAR spoofing attacks have been proposed in existing works.}
(1) \textit{Laser projection}. As shown in Figure \ref{fig:2-spoof-attacks}(b), 
the adversary uses a photodiode to synchronize with the victim LiDAR, and then delays the received laser pulses. 
Then he can choose the fake points that appear in the point cloud by crafting a pulse waveform. Previous works have demonstrated the possibility of relaying LiDAR laser pulses from different locations \cite{spoof-lidar-bhe15} or controlling fake points at different positions in the point cloud \cite{spoof-lidar-shin,jin2022pla,spoof-lidar-ccs19,spoof-lidar-us20} for LiDAR spoofing attacks. (2) \textit{Shape manipulation}. Since each point position in the point cloud also depends on the shapes of the target 3D object, the adversary can craft some objects with carefully-designed shapes to deceive the LiDAR \cite{spoof-lidar-camera-sp21}. (3) \emph{Object placement}. The adversary can use existing objects and place them in identified positions to generate counterfeit laser points and interfere with the perception results of the point cloud model \cite{spoof-lidar-ccs21}.

\noindent\textbf{\revision{1) Spoofer Property.}} 
(1) For the laser projection technique, the spoofer consists of a photodiode, a laser diode, a laser driver module and a delay generator. The prices of the first three devices are about \$1, \$20 and \$150 respectively. The delay generator costs thousands of dollars and its size is as big as a microwave (cost: \$\$\$, size: S2). \revisionB{It spoofs fake points with the laser signal, which is an invisible light and hard to be noticed by the victim (signal: \laser, recog.: False).} 
(2) For the shape manipulation technique, the adversary can use a 3D printer to create a well-designed 3D object that can generate abnormal point clouds (signal: \shape). A 3D printer costs hundreds of dollars (cost: \$\$). The printed adversarial object commonly has unique and noticeable shape in the physical world to effectively fool the point cloud model \cite{spoof-lidar-camera-sp21} (size: S1, recog.: True). (3) For the object placement technique, the adversary places some existing objects to generate extra laser points (signal: \shape). For example, the adversary can control drones to hover around other obstacle's locations \cite{spoof-lidar-ccs21} (cost: \$\$, size S1). The victim passengers can be alert when observing multiple drones constantly flying in front of them (recog.: True).

\noindent\textbf{\revision{2) Spoofing Operation.}} 
The maximum effective attack distance of both laser projection and shape manipulation techniques depends on the firing range of the LiDAR, which is commonly up to 100m (range: R2). 
The attacks aim to fool the point cloud model for object detection. The laser projection technique needs the adversary to actively inject the counterfeit laser pulses (exposure: A) while the other two techniques deceive the LiDAR by placing the adversarial objects without \mbox{runtime intervention (exposure: P).}

\noindent\textbf{\revision{3) Victim Characteristic.}} 
Almost all existing LiDAR spoofing attacks target outdoor autonomous vehicles \cite{spoof-lidar-bhe15,spoof-lidar-shin,jin2022pla,spoof-lidar-ccs19,spoof-lidar-us20,spoof-lidar-camera-sp21} (type: \car, scenario: \outdoor). 

\noindent\textbf{\revision{4) Attack Goal.}} 
All the three techniques aim to tamper with the perceived environment (state: Object) and lead the RV to make wrong actions (\revision{AtkPath: 4}). Specifically, (1) some attacks create a non-existence object (e.g., wall \cite{spoof-lidar-bhe15}, vehicle \cite{spoof-lidar-us20},  arbitrary objects \cite{spoof-lidar-ccs19}) in front of the victim vehicle. This can cause two consequences: the victim vehicle has to perform a hard brake, which could hurt the passengers inside or cause rear-end collisions; if the spoofed object is placed at the cross-road, it could freeze the victim vehicle even the traffic light is green, causing heavy traffic congestion. (2) Some attacks erase existing objects from the victim's perception output \cite{spoof-lidar-shin,spoof-lidar-camera-sp21,spoof-lidar-ccs21}. As a result, the victim vehicle cannot recognize the objects and will crash into them. 

\subsection{Camera Spoofing Attacks}

\xuyuan{
A camera is an optical-electrical device that converts the light perceived by lens to electrical signals.
}
The adversary can alter the results by adding visual perturbations. 
\noindent\textbf{\revision{1) Spoofer Property.}} 
(1) The sticker-pasting technique creates a counterfeit patch to fool the camera (signal: \sticker). Due to the physical constraints, the adversarial patch is still visually abnormal (size: S1, recog.: True). It is very cheap to print such a patch (cost: \$). 
(2) The light-projection technique can be implemented in two ways: using a laser pointer to shine laser beams on the target object or into camera, or using a projector to project adversarial images on the road or wall (signal: \laser). Both spoofers are small and can be mounted on a drone \cite{spoof-camera-light-ccs20,spoof-camera-light-iacr20} (size: S1). A common laser pointer costs hundreds of dollars (cost: \$), which is much cheaper than a projector with high lumen intensity and resolutions (cost: \$\$). 
\revisionB{Unlike LiDAR, camera spoofing attacks can be implemented using either visible natural light  (recog.: True) or invisible infrared light (recog.: False) \cite{spoof-camera-light-ccs21}. In particular, infrared light exploits a portion of the spectrum that is invisible to humans but detectable by cameras.}

\noindent\textbf{\revision{2) Spoofing Operation.}} 
Camera spoofing is commonly launched under the post-processing setting with a large attack range (range: R2). 

\noindent\textbf{\revision{3) Victim Characteristic.}} 
Most works on camera spoofing attacks target the object or lane detection function in the outdoor autonomous driving scenario \cite{spoof-camera-stick-cvpr18,spoof-camera-stick-woot18,spoof-camera-stick-ccs19,kong2020physgan,wang2021dual,spoof-camera-stick-iclr20,spoof-camera-stick-us21,spoof-camera-stick-us21-cq,spoof-camera-raid,yan2022rolling,spoof-camera-light-ccs20,spoof-camera-light-iacr20,spoof-camera-light-cvpr21,spoof-camera-light-ccs21,spoof-camera-light-us21} (type: \car, scenario: \outdoor). One exception is \cite{spoof-camera-light-woot16}, which uses light-projection spoofers to counterfeit the lateral drift velocity and induces the drone to follow the motion to compensate the spoofed drift (type: \drone).

\noindent\textbf{\revision{4) Attack Goal.}} 
There are different goals and attack paths for camera spoofing attacks. \xingshuo{(1) Object detection: a quantity of works attack the traffic controller detection~\cite{spoof-camera-stick-cvpr18,spoof-camera-stick-woot18,spoof-camera-stick-ccs19,kong2020physgan,wang2021dual,zhong2022shadows}. They cause the vehicle to make wrong classification results and control actions, e.g., a stop sign is mis-classified as a speed limit sign. Some works target the obstacle detection function to make the RV detect a non-existence obstacle \cite{spoof-camera-raid,yan2022rolling,spoof-camera-light-ccs20,spoof-camera-light-iacr20,spoof-camera-light-cvpr21,spoof-camera-light-ccs21,spoof-camera-light-us21}, or miss an existing obstacle~\cite{spoof-camera-stick-ccs19,cheng2022physical,huang2020universal,wu2020making,xu2020adversarial,zhai2020s}. 
All these works belong to the category of (state: Object, \revision{AtkPath: 4}).
(2) Lane detection: some works \cite{spoof-camera-stick-us21,spoof-camera-stick-us21-cq,spoof-camera-light-iacr20,spoof-camera-light-ccs20} change the correct trajectory of the vehicle by misleading lane markings on the highway (state: Lane, \revision{AtkPath: 4}). (3) Object tracking: Jia et al. \cite{spoof-camera-stick-iclr20} extended the sticker pasting attack to multiple object tracking tasks, which deceives the RV through continuously spoofing the position of the obstacle (state: Object, \revision{AtkPath: 5}). 
(4) Davidson et al. \cite{spoof-camera-light-woot16} proposed to use the counterfeit lateral drift velocity to destabilize the control system and cause further damages (state: Velocity, \revision{AtkPath: 1}).}

\subsection{IMU Spoofing Attacks}

The Inertial Measurement Unit (IMU) is one core sensor to help RVs adjust the speed of the rotors or motors for stabilizing the balance. It consists of a gyroscope, an accelerometer and a magnetometer to measure the rotation, acceleration, and orientation of an RV. 
The IMU commonly adopts the Micro-electromechanical (MEMS) technology. Specifically, the gyroscope and accelerometer use a similar mass-spring structure. Once the RV moves, the sensing mass will vibrate continuously, which
changes the capacitance and then induces electrical signals. The signals will be digitized by the analog-to-digital converter (ADC) and output linear and angular rates. 
Recent works show that both gyroscope and accelerometer are susceptible to resonant acoustic interference \cite{anlysis-imu-imece07,anlysis-imu-isie07,anlysis-imu-tie11,spoof-imu-us15,spoof-imu-eurosp17,spoof-imu-blackhat17,spoof-imu-asiaccs18,spoof-imu-us18}. As shown in Figure \ref{fig:2-spoof-attacks}(d), the adversary can generate sound waves with the similar frequency as the spring-mass structure. Such acoustic signals can set up resonance, forcing the sensing mass to move and spoofing the designated values.
 
\noindent\textbf{\revision{1) Spoofer Property.}} 
An IMU spoofer consists of a function generator, a sound amplifier and a tweeter speaker. Since the resonant frequency of the victim IMU is commonly below 1MHz \cite{spoof-imu-eurosp17,spoof-imu-us18}, a \$320 low-end function generator with the maximum frequency of 20MHz is effective for generating fake signals \cite{spoof-imu-blackhat17} (cost: \$\$, size: S2, signal: \voice). It needs to be operated near its maximum amplitude (around 110dB Source Pressure Level) \cite{spoof-imu-eurosp17,spoof-imu-us18,spoof-imu-sp21}, which is hard not to be noticed (recog.: True).

\noindent\textbf{\revision{2) Spoofing Operation.}} 
The attack distance of the IMU spoofer depends on the strengths of the received modulated acoustic signal, which is further determined by the signal transmitting power. Son et al. \cite{spoof-imu-us15} showed the attack distance can reach 37.58m with LRAD 450XL \cite{lrad-450xl} or UltraElectronics UyperShield \cite{ultraelectronics-hypershield} (range: R2). The spoofer actively injects resonant noise to the victim sensor (exposure: A), and induce the it to directly generate spoofed raw data. 

\noindent\textbf{\revision{3) Victim Characteristic.}} 
The IMU spoofing attack can cover almost all types of RVs for both indoor and outdoor scenarios (type: \mb \drone \car, scenario: \indoor\,\outdoor).

\noindent\textbf{\revision{4) Attack Goal.}} 
There are two types of goals for IMU spoofing attacks. The first type is to tamper with the linear or angular velocity,
and make the victim RV \revisionC{lose} control with the spoofed velocity (state: Velocity, \revision{AtkPath: 1}). For instance, Trippel et al. \cite{spoof-imu-eurosp17} proposed the \emph{output biasing} and \emph{output control} attacks on the MEMS accelerometer. They were then extended to the gyroscope with the \emph{side-swing} and \emph{switching} attacks \cite{spoof-imu-blackhat17,spoof-imu-us18}. Nashimoto et al. \cite{spoof-imu-asiaccs18} further discussed the attack that involves an accelerometer, gyroscope and magnetometer. The second type is to spoof the sensors to fool the victim RV for object detection (state: Object, \revision{AtkPath: 6}). Ji et al. \cite{spoof-imu-sp21} introduced a novel attack that could deceive the image-stabilization-based objection detection function in the autonomous driving system. They found the spoofed IMU data can cause an object to become undetected, mis-classified or create a non-existent object.

\subsection{Microphone Spoofing Attacks}

The microphone is the key component of a voice-control system for human-RV interaction \cite{spoof-voice-wc19}. For example, the user can issue voice commands to specify the destination for autonomous driving, or launch a mission by a service robot. Commonly, RVs use the MEMS microphone, which consists of a transducer, an amplifier, a low pass filter (LPF), and an ADC. 
When a sound wave is received, the air pressure flexes the membrane in the transducer and changes the capacitance \cite{voice-secon16}. 
The LPF and ADC filter the amplified signals beyond the frequency range of human hearing (20Hz $\sim$ 20kHz). 

The  microphone spoofing attack aims to make the target system execute malicious voice commands without being detected or recognized by normal users. Recent attacks leverage two methods to achieve this goal: (Figure \ref{fig:2-spoof-attacks}(e)). (1) \emph{Inaudible voice} technique: the adversary replays synthetic ultrasound signals and disguises them to legitimate digital speech signals based on the nonlinearity of the amplifier \cite{spoof-voice-inaudible-ccs17,spoof-voice-inaudible-mobisys17,spoof-voice-inaudible-poster-ccs17,spoof-voice-inaudible-nsdi18,spoof-voice-inaudible-ndss20,spoof-voice-inaudible-tdsc21}. (2) \emph{Audio injection} technique: the adversary hides the adversarial audio in the background noise or songs \cite{spoof-voice-audio-us16,spoof-voice-audio-us18,spoof-voice-audio-sp18,spoof-voice-audio-ndss19,spoof-voice-audio-us20}.

\noindent\textbf{\revision{1) Spoofer Property.}} 
The microphone spoofer in the inaudible voice method is similar as the IMU spoofer while the function generator requires a much larger sampling range (cost: \$\$\$, size: S2). The audio injection method can be achieved with a media player (cost: \$, size: S1).
Both methods are difficult to be noticed by the victim since the malicious voice is either inaudible or hidden (signal: \voice, recog.: False).

\noindent\textbf{\revision{2) Spoofing Operation.}} 
Over the years, the effective attack distance for the inaudible voice technique has been increased from 1.75m \cite{spoof-voice-inaudible-ccs17} to 19.8m \cite{spoof-voice-inaudible-tdsc21} (range: R2). The adversary shifts a high-frequency inaudible signal to a low-frequency audible signal, and then injects it to the speech recognition function (exposure: A). In contrast, the audio injection technique aims to hide the malicious acoustic commands into the normal audio waveform (range: R2, exposure: P).

\noindent\textbf{\revision{3) Victim Characteristic.}} 
Recent works on microphone spoofing attacks target the voice-control system in autonomous vehicles \cite{spoof-voice-wc19,spoof-voice-inaudible-tdsc21} (type: \car, scenario: \outdoor). Modern vehicles (e.g., Tesla \cite{tesla-voice-commands}, Audi \cite{audi-voice-commands}, Lincoln \cite{lincoln-voice-commands}) support a list of voice commands that will be converted into the navigation goal, which are vulnerable to the spoofing attacks.

\noindent\textbf{\revision{4) Attack Goal.}} 
The adversary generates malicious voice commands, which are further converted into unexpected missions for the RV to execute (state: mission goal, \revision{AtkPath: 5}). For instance, Yan et al. \cite{spoof-voice-inaudible-tdsc21} applied the inaudible voice technique \cite{spoof-voice-inaudible-ccs17,spoof-voice-inaudible-poster-ccs17,spoof-voice-inaudible-mobisys17,spoof-voice-inaudible-nsdi18,spoof-voice-inaudible-ndss20} to manipulate some in-car features in an Audi autonomous vehicle, such as navigation, entertainment, environmental controls and mobile phone control. Zhou et al. \cite{spoof-voice-wc19} discussed the possibility of using the audio injection method \cite{spoof-voice-audio-us16,spoof-voice-audio-us18,spoof-voice-audio-sp18,spoof-voice-audio-icml19,spoof-voice-audio-us20} to control navigation functions in autonomous driving systems as well.

\subsection{Ultrasonic \& MMW Radar Spoofing Attack}
\label{sec:3-radar}

The ultrasonic sensor and MMW radar also utilize the ToF method to measure the distance between the RV and \revisionC{an} obstacle. The ultrasonic sensor transmits and receives ultrasound waves, which have a relatively low speed (340m/s) and are vulnerable to bad weather. Therefore, it is commonly used in simple scenarios, such as automatic parking. 
In contrast, the MMW radar relies on the millimeter waves, and is widely used in outdoor autonomous vehicles. It can assist LiDAR and cameras to detect obstacles under extreme weather conditions. \xuyuan{In addition, it can also be used to track objects and estimate their velocity.} As shown in Figure \ref{fig:2-spoof-attacks}(f), the adversary can spoof these two sensors by relaying the received signal and sending it back to the transmitter \cite{spoof-radar-phdthesis,spoof-ultrosonic-defcon16,spoof-ultrosonic-iotj18,spoof-ultrosonic-ashes19,spoof-radar-tifs21}. 

\noindent\textbf{\revision{1) Spoofer Property.}} 
The ultrasonic spoofer consists of an envelope detector, ultrasonic transducers, amplification circuits, a buffer amplifier and a square wave generator. A recent work \cite{spoof-ultrosonic-iotj18} showed an Arduino board at the price of \$20 can generate the required square waves of the selected frequency (40$\sim$50kHz) (cost: \$, size: S1, signal: \voice). 
Due to the large frequency range in the radar, e.g., 76$\sim$77GHz MRR Radar installed on Tesla, an effective spoofer that works at such high frequency can cost more than 10 thousand dollars with a large size \cite{spoof-ultrosonic-defcon16} (cost: \$\$\$, size: S2, signal: \wave). Both ultrasound and MMW radar waves are imperceptible (recog.: False).

\noindent\textbf{\revision{2) Spoofing Operation.}} 
Both attacks generate counterfeit signals and actively inject them into the victim's sensors (exposure: A). The attack distance of the ultrasonic spoofing attack is up to 2 meters \cite{spoof-ultrosonic-iotj18} (range: R1) while that of the radar spoofing attack can reach 26m \cite{spoof-radar-tifs21} (range: R2).

\noindent\textbf{\revision{3) Victim Characteristic.}} 
Both attacks target autonomous vehicles in an outdoor scenario (type: \car, scenario: \outdoor). The ultrasonic spoofing attack can also be applied in an indoor parking lot (scenario: \indoor).

\noindent\textbf{\revision{4) Attack Goal.}} 
Existing spoofing attacks aim to create a non-existence object in front of the victim RV or falsify the location of an existing obstacle (state: Object). These attacks lead to two different attack paths. (1) Some works launch the ultrasonic spoofing attack \cite{spoof-ultrosonic-defcon16,spoof-ultrosonic-iotj18} against different commercial autonomous vehicles. Miura et al. \cite{spoof-ultrosonic-ashes19} reduced the cost of radar spoofing attack by using a replica radar and one additional small Micro-Control Unit (MCU). Sun et al. \cite{spoof-radar-tifs21} deployed radar spoofing attacks to cause vehicle stalling, hard braking and lane changing. These attacks follow the \revision{AtkPath4}. (2) Sun et al. \cite{spoof-radar-tifs21} also proposed the multi-stage and cruise control attacks, which lead to a high speed crash by leveraging a long-term plan and control in autonomous driving. This is \revision{AtkPath5}.

\begin{figure*}[tb]
\centering
\includegraphics[width=1.00\linewidth]{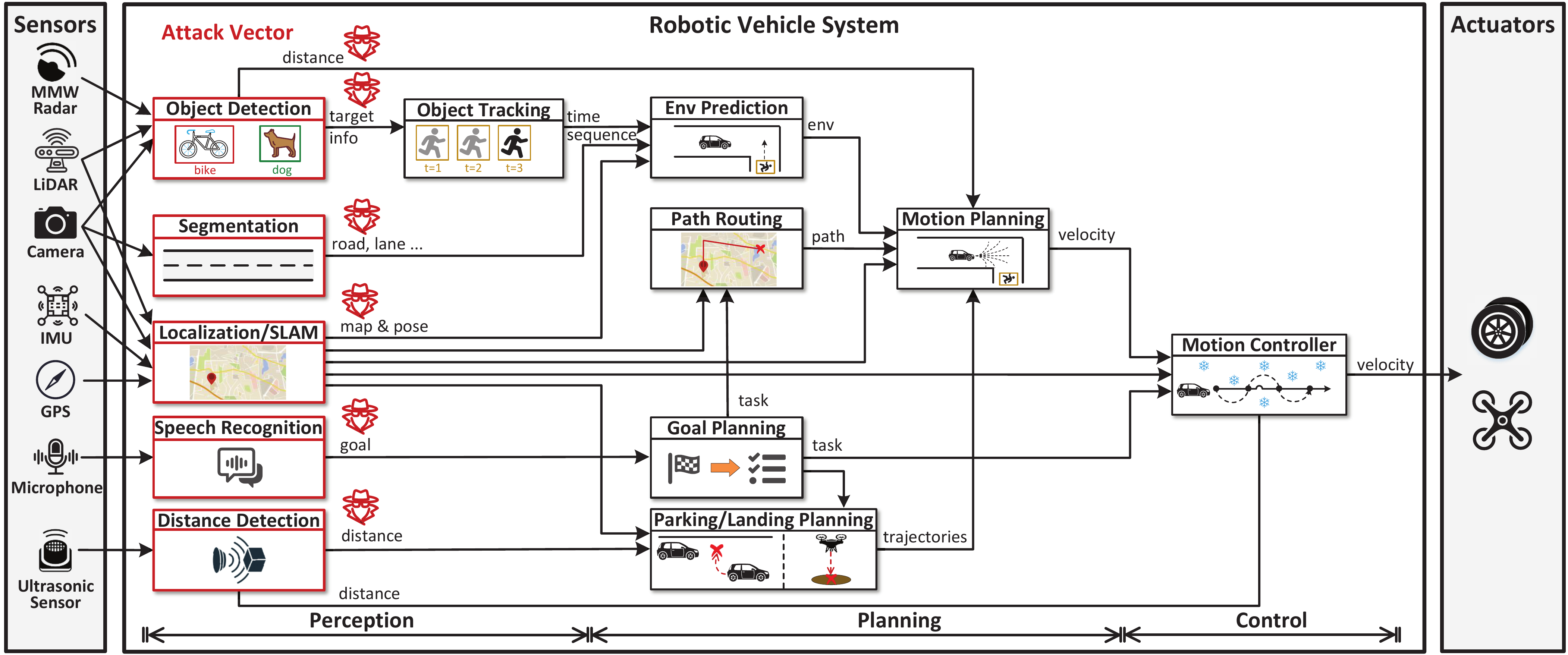}
\vspace{-15pt}
\caption{The action flow model of RV systems.}
\vspace{-15pt}
\label{fig:3-workload-model}
\end{figure*}

\subsection{Insights and Lessons}
\label{sec:2-threat-model}

We identify some interesting observations and lessons from the above systematization. 

\begin{tcolorbox}[left=1mm, right=1mm, top=0.5mm, bottom=0.5mm, arc=1mm]
\textbf{Insight 1:} \textit{Compared to autonomous vehicles, there are relatively fewer attacks targeting drones and automated guided vehicles.}
\end{tcolorbox}
\vspace{-5pt}
\revisionS{According to Table \ref{table:spoofing-compare}, we can observe that the majority of works (38/48) target autonomous vehicles, which share widely-considered attack goals and scenarios.} 
Since all the three types of RVs share many functions and mechanisms, those spoofing attacks might be applied to the drones and automated guided vehicles as well. Considering they are playing more important roles in our daily life, such as delivery, photography, surveillance and house cleaning, more efforts should be devoted to the study of attacks against them. \revisionS{Their unique features (e.g., limited computing resources, low-end sensors, eyes in the sky) also lead to other types of attacks, such as cloud-based DoS attacks \cite{rv-dos-atk} and privacy invasion attack \cite{uav-cryp-sp19}. }

\begin{tcolorbox}[left=1mm, right=1mm, top=0.5mm, bottom=0.5mm, arc=1mm]
\textbf{Insight 2:} \textit{The same attack path and consequence can be triggered by different sensors and spoofers, thus giving the adversary more choices to conduct attacks.}
\end{tcolorbox}
\vspace{-5pt}
As shown in Figure \ref{fig:2-atk-model}, each attack path starts from the mis-estimated states (\one/\two) and ends with \revision{abnormal actions} (\ten). So once the operation or environment state is compromised, the subsequent wrong executions will occur correspondingly. 
\revision{The goal of the attack on RVs is to spoof the RV's state and ultimately cause it to perform dangerous actions. Since some sensors provide similar functionalities to RVs (e.g., LiDAR and camera), attackers can achieve the same attack goal through different attack paths. It is important to note that even though the attack results may be the same, the attack methods can be different depending on the sensor being used for state estimation. This insight can inspire us to analyze spoofing attacks that go from sensor to actuator and thus} \revisionS{identify more unexplored attacks.} 
For instance, the SLAM function enables an RV to localize itself with the camera or LiDAR inputs. Then the position-dependent \revision{AtkPath2} and \revision{AtkPath3} related to the GPS spoofers in Table \ref{table:spoofing-compare} can also be implemented by the camera or LiDAR spoofers. The adversary can choose the most convenient and efficient attack solution to achieve his desired goal. 

\begin{tcolorbox}[left=1mm, right=1mm, top=0.5mm, bottom=0.5mm, arc=1mm]
\textbf{Insight 3:} \textit{There are much fewer studies about the multi-round spoofing attacks.}
\end{tcolorbox}
\vspace{-5pt}
Among all the seven attack paths, \revision{AtkPath6} and \revision{AtkPath7} are multi-round attacks, which are the combinations of two single-round paths: the adversary first exploits an $x_t^r$-related attack path to change RV's perceived environmental measurements, and then causes the spoofing attack on an $x_t^e$-related path. The interaction between the RV and its surrounding environment bridges these two distinct attack paths. From Table \ref{table:spoofing-compare}, we observe there are only 3 out of 42 works exploring two specific multi-round attacks: target blurring and ROI altering attacks. We hypothesize that the lack of such studies is due to the complexity of multi-round mechanisms and RV-environment interactions. We believe there are more opportunities for designing such attacks. On one hand, researchers can use other sensor spoofers to implement the target blurring and ROI altering attacks; on the other hand, more multi-round attack paths besides \revision{AtkPath6} and \revision{AtkPath7} can be explored in the future. 

\begin{tcolorbox}[left=1mm, right=1mm, top=0.5mm, bottom=0.5mm, arc=1mm]
\textbf{Insight 4:} \textit{GPS, Microphone, MMW Radar spoofing and laser projection are more stealthy than others.}
\end{tcolorbox}
\vspace{-5pt}
A successful spoofing attack should be stealthy enough to avoid the detection of the victim. To achieve this, we identify the following combinations of spoofer characteristics. (1) \emph{Passive exposure} + \emph{small size}: a passive attack is to hide the spoofer in the environment without human intervention,  so it is hard for the victim to recognize the existence of the spoofer. Thus, the spoofer is imperceptible as long as it is small enough and highly integrated into the environment, e.g., the audio injection attack. 
(2) \emph{Active exposure} + \emph{imperceptible signal} + \emph{remote attack range}: 
\revision{Spoofing at a long distance (e.g. 10 meters) is stealthy even if someone is observing, because spoofers are too small to be identified at such a long distance. In addition, the imperceptible signal (e.g., GPS, laser, ultrasonic and MMW radar) allows the victim to detect the attack only by observing the cheater at a distance, which further ensures the stealthiness.}
We encourage that other spoofing attacks can improve their stealthiness from these two dimensions.
\section{A Unified Model for Spoofing Attacks}
\label{sec:3}
We propose a novel action flow model to systematically describe the spoofing attacks.  
This model not only covers all the existing attacks summarized in \cref{sec:review}, but also reveals new attacks that are never considered previously. 

\subsection{Action Flow Model}
\label{sec:3-af-model}
Existing studies on sensor spoofing attacks tend to follow an ad-hoc way: researchers first identify the mis-estimated state that can incur the desired consequence, and then design the spoofing methodology that can falsify the state. 
The lack of systematization cannot guarantee the comprehensiveness of the discovered \revision{unexplored sensor spoofing threats}. To address this limitation, we build an \textit{action flow model} to abstract possible control flows in different RV systems and scenarios. Each control flow can lead to some potential spoofing attacks. 
\revision{It is important to note that the action flow and attack path are different: action flow only describes an RV system while attack path is an extension of action flow with security considerations. More specifically, attack path is a higher-level abstraction of RV systems, helping distinguish the target states of each sensor spoofing attack. Action flow is specific to each sensor type and the function type inside the system, helping identify new unexplored attacks.}

Figure \ref{fig:3-workload-model} shows our action flow model, which consists of 12 interacted robot functions. This is established by surveying the function compositions and interactions of various RVs from both industrial ecosystems \cite{rv-system-apollo,rv-system-dji} and academic papers \cite{mavbench,xuyuan-raid,gelei-raid}. Among them, sensors directly interact with five functions: \textit{Objection Detection}, \textit{Segmentation}, \textit{Localization/SLAM}, \textit{Speech Recognition} and \textit{Distance Detection}. Then these functions further interact with other subsequent functions. First, \textit{Objection Detection} recognizes nearby obstacles in real-time. Then \textit{Object Tracking} assigns each predicted obstacle an ID and dynamically tracks it. Meanwhile, \textit{Segmentation} estimates the static environmental information. 
Combining these two, \textit{Environment Prediction} estimates the obstacle trajectories with probabilities/priorities. Second, \textit{Localization/SLAM} estimates the operation states of the RV based on the known/unknown map. This information is the key to judge which obstacles in the surrounding environment have a high possibility of interacting with the RV. Third, \textit{Speech Recognition} provides an interface for executing new tasks through human-computer interaction. \textit{Goal Planning} transfers each task goal to a set of sub-tasks and dispatches them to other planning functions. For example, \textit{Path Routing} calculates a path from the estimated position to the destination given by \textit{Goal Planning}. \textit{Parking/Landing Planning} plans a set of trajectories to complete the parking or landing task. Both the planned path and the trajectory will be sent to \textit{Motion Planning}, which outputs the suitable velocity to avoid obstacles. Finally, \textit{Motion Controller} transfers this velocity to each actuator and ensures the stability of the RV.

\setlength{\tabcolsep}{1pt}
\begin{table*}[tbh]
\vspace{-10pt}
\caption{Possible attacks in different flow patterns.}
\vspace{-5pt}
\centering
\arrayrulecolor{black}
\resizebox{1.0\linewidth}{!}{
\begin{tabular}{|c|l|cc|cccc|ccc|cc|cc|cc|cc|}
\rowcolor{gray!60}
\hline
\cellcolor{gray!60} & \cellcolor{gray!60} & \multicolumn{2}{c|}{\cellcolor{gray!60}} & \multicolumn{4}{c|}{\cellcolor{gray!60}\textbf{LiDAR}} & \multicolumn{3}{c|}{\cellcolor{gray!60}\textbf{Camera}} & \multicolumn{2}{c|}{\cellcolor{gray!60}} & \multicolumn{2}{c|}{\cellcolor{gray!60}} & \multicolumn{2}{c|}{\cellcolor{gray!60}} & \multicolumn{2}{c|}{\cellcolor{gray!60}} \\
\hhline{|>{\arrayrulecolor{gray!60}}---->{\arrayrulecolor{black}}------->{\arrayrulecolor{gray!60}}-------->{\arrayrulecolor{black}}|} 
\rowcolor{gray!60}
\multirow{-2}{*}{\cellcolor{gray!60}\textbf{\begin{tabular}[c]{@{}c@{}}Flow\\Pattern\end{tabular}}} & \multicolumn{1}{c|}{\multirow{-2}{*}{\cellcolor{gray!60}\textbf{\begin{tabular}[c]{@{}c@{}}Possible Attacks\\(Target Function)\end{tabular}}}} & \multicolumn{2}{c|}{\multirow{-2}{*}{\cellcolor{gray!60}\textbf{MMW Radar}}} & \multicolumn{1}{c|}{\cellcolor{gray!60}\textbf{LP}} & \multicolumn{1}{c|}{\cellcolor{gray!60}\textbf{SM}} & \multicolumn{1}{c|}{\cellcolor{gray!60}\textbf{OP}} & \multicolumn{1}{c|}{\cellcolor{gray!60}} & \multicolumn{1}{c|}{\cellcolor{gray!60}\textbf{SP}} & \multicolumn{1}{c|}{\cellcolor{gray!60}\textbf{LP}} & \multicolumn{1}{c|}{\cellcolor{gray!60}} & \multicolumn{2}{c|}{\multirow{-2}{*}{\cellcolor{gray!60}\textbf{GPS}}} & \multicolumn{2}{c|}{\multirow{-2}{*}{\cellcolor{gray!60}\textbf{IMU}}} & \multicolumn{2}{c|}{\multirow{-2}{*}{\cellcolor{gray!60}\textbf{\begin{tabular}[c]{@{}c@{}}Ultrasonic\\Sensor\end{tabular}}}} & \multicolumn{2}{c|}{\multirow{-2}{*}{\cellcolor{gray!60}\textbf{Microphone}}} \\ 
\hline
 \cellcolor{DarkSeaGreen!40} & \cellcolor{DarkSeaGreen!40}\textbf{Obstacle Appearing (A1)} & \multicolumn{1}{c|}{\cite{spoof-ultrosonic-defcon16,spoof-ultrosonic-ashes19,spoof-radar-tifs21}} & \cellcolor{DarkOrange!40} & \multicolumn{1}{c|}{\cite{spoof-lidar-bhe15,spoof-lidar-shin,jin2022pla,spoof-lidar-ccs19,spoof-lidar-us20}} & \multicolumn{1}{c|}{\gatk} & \multicolumn{1}{c|}{\textbf{\gatk}} & \cellcolor{DarkOrange!40} & \multicolumn{1}{c|}{\textbf{\gatk}} & \multicolumn{1}{c|}{\cite{spoof-camera-light-ccs20,spoof-camera-light-iacr20,spoof-camera-light-ccs21,spoof-camera-uav-us22}} & \cellcolor{DarkOrange!40} & \cellcolor{LightGray!40} & \multicolumn{1}{c|}{\cellcolor{LightGray!40}} & \cellcolor{LightGray!40} & \cellcolor{LightGray!40} & \cellcolor{LightGray!40} & \cellcolor{LightGray!40} & \cellcolor{LightGray!40} & \cellcolor{LightGray!40} \\ 

 \hhline{>{\arrayrulecolor{DarkSeaGreen!40}}->{\arrayrulecolor{black}}-->{\arrayrulecolor{DarkOrange!40}}->{\arrayrulecolor{black}}--->{\arrayrulecolor{DarkOrange!40}}->{\arrayrulecolor{black}}-->{\arrayrulecolor{DarkOrange!40}}->{\arrayrulecolor{LightGray!40}}-------->{\arrayrulecolor{black}}}
 \cellcolor{DarkSeaGreen!40} & \cellcolor{DarkSeaGreen!40}\textbf{Obstacle Missing (A1)} & \multicolumn{1}{c|}{\batk} & \cellcolor{DarkOrange!40} & \multicolumn{1}{c|}{\cite{jin2022pla}} & \multicolumn{1}{c|}{\cite{spoof-lidar-camera-sp21}} & \multicolumn{1}{c|}{\cite{spoof-lidar-ccs21}} & \cellcolor{DarkOrange!40} & \multicolumn{1}{c|}{\cite{spoof-camera-stick-ccs19,cheng2022physical,huang2020universal,wu2020making,xu2020adversarial}} & \multicolumn{1}{c|}{\cite{spoof-camera-raid,spoof-camera-light-us21}} & \cellcolor{DarkOrange!40} & \cellcolor{LightGray!40} & \multicolumn{1}{c|}{\cellcolor{LightGray!40}} & \cellcolor{LightGray!40} & \cellcolor{LightGray!40} & \cellcolor{LightGray!40} & \cellcolor{LightGray!40} & \cellcolor{LightGray!40} & \cellcolor{LightGray!40} \\ 

 \hhline{>{\arrayrulecolor{DarkSeaGreen!40}}->{\arrayrulecolor{black}}-->{\arrayrulecolor{DarkOrange!40}}->{\arrayrulecolor{black}}--->{\arrayrulecolor{DarkOrange!40}}->{\arrayrulecolor{black}}-->{\arrayrulecolor{DarkOrange!40}}->{\arrayrulecolor{LightGray!40}}-------->{\arrayrulecolor{black}}}
 \multirow{-3}{*}{\cellcolor{DarkSeaGreen!40}\textbf{FP1}}& \cellcolor{DarkSeaGreen!40}\textbf{\begin{tabular}[l]{@{}l@{}}Traffic Controller\\Misclassification (A1)\end{tabular}} & \multicolumn{1}{c|}{\cellcolor{LightGray!40}} & \multirow{-3}{*}{\cellcolor{DarkOrange!40}\begin{tabular}[c]{@{}c@{}}AF1\\{[\car]}\end{tabular}} & \multicolumn{1}{c|}{\textbf{\batk}} & \multicolumn{1}{c|}{\batk} & \multicolumn{1}{c|}{\textbf{\batk}} & \multirow{-3}{*}{\cellcolor{DarkOrange!40}\begin{tabular}[c]{@{}c@{}}AF3\\{[\mb\,\car]}\end{tabular}} & \multicolumn{1}{c|}{\cite{spoof-camera-stick-cvpr18,spoof-camera-stick-woot18,spoof-camera-stick-ccs19,kong2020physgan,wang2021dual}} & \multicolumn{1}{c|}{\cite{spoof-camera-raid,yan2022rolling,spoof-camera-light-ccs20,spoof-camera-light-iacr20,spoof-camera-light-cvpr21,spoof-camera-light-ccs21}} & \multirow{-3}{*}{\cellcolor{DarkOrange!40}\begin{tabular}[c]{@{}c@{}}AF10\\{[\mb\drone\car]}\end{tabular}} & \multicolumn{2}{c|}{\cellcolor{LightGray!40}} & \multicolumn{2}{c|}{\cellcolor{LightGray!40}} & \multicolumn{2}{c|}{\cellcolor{LightGray!40}} & \multicolumn{2}{c|}{\cellcolor{LightGray!40}} \\ 
\hhline{*{11}{-}>{\arrayrulecolor{LightGray!40}}-------->{\arrayrulecolor{black}}}
 \cellcolor{DarkSeaGreen!40} & \cellcolor{DarkSeaGreen!40}\textbf{Trajectory Appearing (B)} & \multicolumn{1}{c|}{\cite{spoof-radar-tifs21}} & \cellcolor{DarkOrange!40} & \multicolumn{1}{c|}{\ratk} & \multicolumn{1}{c|}{\ratk} & \multicolumn{1}{c|}{\ratk} & \cellcolor{DarkOrange!40} & \multicolumn{1}{c|}{\cite{spoof-camera-stick-iclr20}} & \multicolumn{1}{c|}{\gatk} & \cellcolor{DarkOrange!40} & \cellcolor{LightGray!40} & \multicolumn{1}{c|}{\cellcolor{LightGray!40}} & \cellcolor{LightGray!40} & \cellcolor{LightGray!40} & \cellcolor{LightGray!40} & \cellcolor{LightGray!40} & \cellcolor{LightGray!40} & \cellcolor{LightGray!40} \\ 

 \hhline{>{\arrayrulecolor{DarkSeaGreen!40}}->{\arrayrulecolor{black}}-->{\arrayrulecolor{DarkOrange!40}}->{\arrayrulecolor{black}}--->{\arrayrulecolor{DarkOrange!40}}->{\arrayrulecolor{black}}-->{\arrayrulecolor{DarkOrange!40}}->{\arrayrulecolor{LightGray!40}}-------->{\arrayrulecolor{black}}}
 \cellcolor{DarkSeaGreen!40} & \cellcolor{DarkSeaGreen!40}\textbf{Trajectory Missing (B)} & \multicolumn{1}{c|}{\batk} & \cellcolor{DarkOrange!40} & \multicolumn{1}{c|}{\ratk} & \multicolumn{1}{c|}{\ratk} & \multicolumn{1}{c|}{\ratk} & \cellcolor{DarkOrange!40} & \multicolumn{1}{c|}{\cite{spoof-camera-stick-iclr20}} & \multicolumn{1}{c|}{\gatk} & \cellcolor{DarkOrange!40} & \cellcolor{LightGray!40} & \multicolumn{1}{c|}{\cellcolor{LightGray!40}} & \cellcolor{LightGray!40} & \cellcolor{LightGray!40} & \cellcolor{LightGray!40} & \cellcolor{LightGray!40} & \cellcolor{LightGray!40} & \cellcolor{LightGray!40} \\  

 \hhline{>{\arrayrulecolor{DarkSeaGreen!40}}->{\arrayrulecolor{black}}-->{\arrayrulecolor{DarkOrange!40}}->{\arrayrulecolor{black}}--->{\arrayrulecolor{DarkOrange!40}}->{\arrayrulecolor{black}}-->{\arrayrulecolor{DarkOrange!40}}->{\arrayrulecolor{LightGray!40}}-------->{\arrayrulecolor{black}}}
\multirow{-3}{*}{\cellcolor{DarkSeaGreen!40}\textbf{FP2}} & \cellcolor{DarkSeaGreen!40}\textbf{\begin{tabular}[l]{@{}l@{}}Trajectory Altering (C1)\end{tabular}} & \multicolumn{1}{c|}{\batk} & \multirow{-3}{*}{\cellcolor{DarkOrange!40}\begin{tabular}[c]{@{}c@{}}AF2\\{[\car]}\end{tabular}} & \multicolumn{1}{c|}{\ratk} & \multicolumn{1}{c|}{\ratk} & \multicolumn{1}{c|}{\textbf{\ratk}} & \multirow{-3}{*}{\cellcolor{DarkOrange!40}\begin{tabular}[c]{@{}c@{}}AF4\\{[\mb\,\car]}\end{tabular}} & \multicolumn{1}{c|}{\cite{spoof-camera-stick-iclr20}} & \multicolumn{1}{c|}{\gatk} & \multirow{-3}{*}{\cellcolor{DarkOrange!40}\begin{tabular}[c]{@{}c@{}}AF11\\{[\mb\,\car]}\end{tabular}} & \multicolumn{2}{c|}{\cellcolor{LightGray!40}} & \multicolumn{2}{c|}{\cellcolor{LightGray!40}} & \multicolumn{2}{c|}{\cellcolor{LightGray!40}} & \multicolumn{2}{c|}{\cellcolor{LightGray!40}} \\ 
\hhline{*{11}{-}>{\arrayrulecolor{LightGray!40}}-------->{\arrayrulecolor{black}}}
\cellcolor{DarkSeaGreen!40}\textbf{FP3} & \multicolumn{1}{l|}{\cellcolor{DarkSeaGreen!40}\textbf{Lane Altering (A2)}} & \multicolumn{2}{c|}{\cellcolor{LightGray!40}} & \multicolumn{4}{c|}{\cellcolor{LightGray!40}} & \multicolumn{1}{c|}{\cite{spoof-camera-stick-us21,spoof-camera-stick-us21-cq}} & \multicolumn{1}{c|}{\cite{spoof-camera-light-iacr20}} & \cellcolor{DarkOrange!40}\begin{tabular}[c]{@{}c@{}}AF12\\{[\mb\,\car]}\end{tabular} & \multicolumn{2}{c|}{\multirow{-7}{*}{\cellcolor{LightGray!40}}} & \multicolumn{2}{c|}{\multirow{-7}{*}{\cellcolor{LightGray!40}}} & \multicolumn{2}{c|}{\cellcolor{LightGray!40}} & \multicolumn{2}{c|}{\cellcolor{LightGray!40}}  \\ 
\hhline{-->{\arrayrulecolor{LightGray!40}}-->{\arrayrulecolor{black}}*{11}{-}>{\arrayrulecolor{LightGray!40}}---->{\arrayrulecolor{black}}}
\cellcolor{DarkSeaGreen!40}\textbf{FP4} & \multicolumn{1}{l|}{\cellcolor{DarkSeaGreen!40}\textbf{\begin{tabular}[l]{@{}l@{}}Deviating Position\\Altering (E)\end{tabular}}} & \multicolumn{2}{c|}{\cellcolor{LightGray!40}} & \multicolumn{1}{c|}{\textbf{\ratk}} & \multicolumn{1}{c|}{\textbf{\ratk}} & \multicolumn{1}{c|}{\textbf{\ratk}} & \cellcolor{DarkOrange!40}\begin{tabular}[c]{@{}c@{}}AF8\\{[\mb\,\car]}\end{tabular} & \multicolumn{1}{c|}{\textbf{\ratk}} & \multicolumn{1}{c|}{\textbf{\ratk}} & \cellcolor{DarkOrange!40}\begin{tabular}[c]{@{}c@{}}AF16\\{[\mb\,\car]}\end{tabular} & \multicolumn{1}{c|}{\cite{spoof-gps-car-us20,spoof-gps-uav-us22}} & \cellcolor{DarkOrange!40}\begin{tabular}[c]{@{}c@{}}AF21\\{[\drone\car]}\end{tabular} & \multicolumn{1}{c|}{\textbf{\batk}} & \cellcolor{DarkOrange!40}\begin{tabular}[c]{@{}c@{}}AF26\\{[\mb\,\car]}\end{tabular} & \multicolumn{2}{c|}{\cellcolor{LightGray!40}} & \multicolumn{2}{c|}{\cellcolor{LightGray!40}} \\ 
\hhline{-->{\arrayrulecolor{LightGray!40}}-->{\arrayrulecolor{black}}*{11}{-}>{\arrayrulecolor{LightGray!40}}---->{\arrayrulecolor{black}}}
\cellcolor{DarkSeaGreen!40}\textbf{FP5} & \multicolumn{1}{l|}{\cellcolor{DarkSeaGreen!40}\textbf{\begin{tabular}[l]{@{}l@{}}Predicted Priority\\Altering (C1)\end{tabular}}} & \multicolumn{2}{c|}{\cellcolor{LightGray!40}} & \multicolumn{1}{c|}{\textbf{\ratk}} & \multicolumn{1}{c|}{\textbf{\ratk}} & \multicolumn{1}{c|}{\textbf{\ratk}} & \cellcolor{DarkOrange!40}\begin{tabular}[c]{@{}c@{}}AF5\\{[\mb\,\car]}\end{tabular} & \multicolumn{1}{c|}{\textbf{\ratk}} & \multicolumn{1}{c|}{\textbf{\ratk}} & \cellcolor{DarkOrange!40}\begin{tabular}[c]{@{}c@{}}AF13\\{[\mb\,\car]}\end{tabular} & \multicolumn{1}{c|}{\textbf{\batk}} & \cellcolor{DarkOrange!40}\begin{tabular}[c]{@{}c@{}}AF18\\{[\car]}\end{tabular} & \multicolumn{1}{c|}{\textbf{\batk}} & \cellcolor{DarkOrange!40}\begin{tabular}[c]{@{}c@{}}AF23\\{[\mb\,\car]}\end{tabular} & \multicolumn{2}{c|}{\cellcolor{LightGray!40}} & \multicolumn{2}{c|}{\cellcolor{LightGray!40}} \\ 
\hhline{-->{\arrayrulecolor{LightGray!40}}-->{\arrayrulecolor{black}}*{11}{-}>{\arrayrulecolor{LightGray!40}}---->{\arrayrulecolor{black}}}
\cellcolor{DarkSeaGreen!40} & \multicolumn{1}{l|}{\cellcolor{DarkSeaGreen!40}\textbf{\begin{tabular}[l]{@{}l@{}}Target Deviating\\Position Altering (D1)\end{tabular}}} & \multicolumn{2}{c|}{\cellcolor{LightGray!40}} & \multicolumn{1}{c|}{\textbf{\ratk}} & \multicolumn{1}{c|}{\textbf{\ratk}} & \multicolumn{1}{c|}{\textbf{\ratk}} & \cellcolor{DarkOrange!40} & \multicolumn{1}{c|}{\textbf{\ratk}} & \multicolumn{1}{c|}{\textbf{\ratk}} & \cellcolor{DarkOrange!40} & \multicolumn{1}{c|}{\cite{spoof-gps-car-us18,spoof-gps-uav-us22}} & \cellcolor{DarkOrange!40} & \multicolumn{1}{c|}{\textbf{\batk}} & \cellcolor{DarkOrange!40} & \multicolumn{2}{c|}{\cellcolor{LightGray!40}} & \multicolumn{2}{c|}{\cellcolor{LightGray!40}}\\ 

\hhline{>{\arrayrulecolor{DarkSeaGreen!40}}->{\arrayrulecolor{black}}->{\arrayrulecolor{LightGray!40}}-->{\arrayrulecolor{black}}--->{\arrayrulecolor{DarkOrange!40}}->{\arrayrulecolor{black}}-->{\arrayrulecolor{DarkOrange!40}}->{\arrayrulecolor{black}}->{\arrayrulecolor{DarkOrange!40}}->{\arrayrulecolor{black}}->{\arrayrulecolor{DarkOrange!40}}->{\arrayrulecolor{LightGray!40}}---->{\arrayrulecolor{black}}}
\multirow{-3}{*}{\cellcolor{DarkSeaGreen!40}\textbf{FP6}} & \multicolumn{1}{l|}{\cellcolor{DarkSeaGreen!40}\textbf{\begin{tabular}[l]{@{}l@{}}Loop Closure\\Failure (A3)\end{tabular}}} & \multicolumn{2}{c|}{\cellcolor{LightGray!40}} & \multicolumn{1}{c|}{\textbf{\batk}} & \multicolumn{1}{c|}{\textbf{\batk}} & \multicolumn{1}{c|}{\textbf{\batk}} & \multirow{-3}{*}{\cellcolor{DarkOrange!40}\begin{tabular}[c]{@{}c@{}}AF6\\{[\mb\,\car]}\end{tabular}} & \multicolumn{1}{c|}{\textbf{\batk}} & \multicolumn{1}{c|}{\textbf{\batk}} & \multirow{-3}{*}{\cellcolor{DarkOrange!40}\begin{tabular}[c]{@{}c@{}}AF14\\{[\mb\,\car]}\end{tabular}} & \multicolumn{1}{c|}{\textbf{\batk}} & \multirow{-3}{*}{\cellcolor{DarkOrange!40}\begin{tabular}[c]{@{}c@{}}AF19\\{[\drone\car]}\end{tabular}} & \multicolumn{1}{c|}{\textbf{\batk}} & \multirow{-3}{*}{\cellcolor{DarkOrange!40}\begin{tabular}[c]{@{}c@{}}AF24\\{[\mb\,\car]}\end{tabular}} & \multicolumn{2}{c|}{\cellcolor{LightGray!40}} & \multicolumn{2}{c|}{\cellcolor{LightGray!40}} \\
\hhline{-->{\arrayrulecolor{LightGray!40}}-->{\arrayrulecolor{black}}*{11}{-}>{\arrayrulecolor{LightGray!40}}---->{\arrayrulecolor{black}}}
\cellcolor{DarkSeaGreen!40} & \multicolumn{1}{l|}{\cellcolor{DarkSeaGreen!40}\textbf{\begin{tabular}[l]{@{}l@{}}Destabilizing Velocity\\Altering (F)\end{tabular}}} & \multicolumn{2}{c|}{\cellcolor{LightGray!40}} & \multicolumn{1}{c|}{\textbf{\batk}} & \multicolumn{1}{c|}{\textbf{\batk}} & \multicolumn{1}{c|}{\textbf{\batk}} & \cellcolor{DarkOrange!40} & \multicolumn{1}{c|}{\textbf{\batk}} & \multicolumn{1}{c|}{\textbf{\batk}} & \cellcolor{DarkOrange!40} & \multicolumn{1}{c|}{\textbf{\batk}} & \cellcolor{DarkOrange!40} & \multicolumn{1}{c|}{\cite{spoof-imu-eurosp17,spoof-imu-blackhat17,spoof-imu-asiaccs18,spoof-imu-us18}} & \cellcolor{DarkOrange!40} & \multicolumn{2}{c|}{\cellcolor{LightGray!40}} & \multicolumn{2}{c|}{\cellcolor{LightGray!40}}\\ 

\hhline{>{\arrayrulecolor{DarkSeaGreen!40}}->{\arrayrulecolor{black}}->{\arrayrulecolor{LightGray!40}}-->{\arrayrulecolor{black}}--->{\arrayrulecolor{DarkOrange!40}}->{\arrayrulecolor{black}}-->{\arrayrulecolor{DarkOrange!40}}->{\arrayrulecolor{black}}->{\arrayrulecolor{DarkOrange!40}}->{\arrayrulecolor{black}}->{\arrayrulecolor{DarkOrange!40}}->{\arrayrulecolor{LightGray!40}}---->{\arrayrulecolor{black}}}
\multirow{-3}{*}{\cellcolor{DarkSeaGreen!40}\textbf{FP7}} & \multicolumn{1}{l|}{\cellcolor{DarkSeaGreen!40}\textbf{\begin{tabular}[l]{@{}l@{}}Destabilizing Position\\Altering (F)\end{tabular}}} & \multicolumn{2}{c|}{\cellcolor{LightGray!40}} & \multicolumn{1}{c|}{\textbf{\batk}} & \multicolumn{1}{c|}{\textbf{\batk}} & \multicolumn{1}{c|}{\textbf{\batk}} & \multirow{-3}{*}{\cellcolor{DarkOrange!40}\begin{tabular}[c]{@{}c@{}}AF9\\{[\mb\,\car]}\end{tabular}} & \multicolumn{1}{c|}{\textbf{\gatk}} & \multicolumn{1}{c|}{\cite{spoof-camera-light-woot16}} & \multirow{-3}{*}{\cellcolor{DarkOrange!40}\begin{tabular}[c]{@{}c@{}}AF17\\{[\mb\,\car\,\drone]}\end{tabular}} & \multicolumn{1}{c|}{\textbf{\batk}} & \multirow{-3}{*}{\cellcolor{DarkOrange!40}\begin{tabular}[c]{@{}c@{}}AF22\\{[\car\,\drone]}\end{tabular}} & \multicolumn{1}{c|}{\textbf{\batk}} & \multirow{-3}{*}{\cellcolor{DarkOrange!40}\begin{tabular}[c]{@{}c@{}}AF27\\{[\mb\,\car\,\drone]}\end{tabular}} & \multicolumn{2}{c|}{\cellcolor{LightGray!40}} & \multicolumn{2}{c|}{\cellcolor{LightGray!40}} \\
\hhline{-->{\arrayrulecolor{LightGray!40}}-->{\arrayrulecolor{black}}*{11}{-}>{\arrayrulecolor{LightGray!40}}---->{\arrayrulecolor{black}}}
\cellcolor{DarkSeaGreen!40}\textbf{FP8} & \multicolumn{1}{l|}{\cellcolor{DarkSeaGreen!40}\textbf{\begin{tabular}[l]{@{}l@{}}Specific Location\\Altering (D2)\end{tabular}}} & \multicolumn{2}{c|}{\cellcolor{LightGray!40}} & \multicolumn{1}{c|}{\textbf{\ratk}} & \multicolumn{1}{c|}{\textbf{\ratk}} & \multicolumn{1}{c|}{\textbf{\ratk}} & \cellcolor{DarkOrange!40}\begin{tabular}[c]{@{}c@{}}AF7\\{[\drone]}\end{tabular} & \multicolumn{1}{c|}{\textbf{\ratk}} & \multicolumn{1}{c|}{\textbf{\ratk}} & \cellcolor{DarkOrange!40}\begin{tabular}[c]{@{}c@{}}AF15\\{[\drone]}\end{tabular} & \multicolumn{1}{c|}{\cite{spoof-gps-drone-vlsid18,spoof-gps-drone-netw19-1,spoof-gps-drone-netw19-2}} & \cellcolor{DarkOrange!40}\begin{tabular}[c]{@{}c@{}}AF20\\{[\drone]}\end{tabular} & \multicolumn{1}{c|}{\textbf{\batk}} & \cellcolor{DarkOrange!40}\begin{tabular}[c]{@{}c@{}}AF25\\{[\drone]}\end{tabular} & \multicolumn{2}{c|}{\cellcolor{LightGray!40}} & \multicolumn{2}{c|}{\cellcolor{LightGray!40}} \\ 
\hhline{-->{\arrayrulecolor{LightGray!40}}-->{\arrayrulecolor{black}}*{13}{-}>{\arrayrulecolor{LightGray!40}}-->{\arrayrulecolor{black}}}
\cellcolor{DarkSeaGreen!40}\textbf{FP9} & \multicolumn{1}{l|}{\cellcolor{DarkSeaGreen!40}\textbf{\begin{tabular}[l]{@{}l@{}}Obstacle Distance\\Altering (A5)\end{tabular}}} & \multicolumn{2}{c|}{\cellcolor{LightGray!40}} & \multicolumn{4}{c|}{\cellcolor{LightGray!40}} & \multicolumn{3}{c|}{\cellcolor{LightGray!40}} & \multicolumn{2}{c|}{\cellcolor{LightGray!40}} & \multicolumn{2}{c|}{\cellcolor{LightGray!40}} & \multicolumn{1}{c|}{\cite{spoof-ultrosonic-defcon16,spoof-ultrosonic-iotj18}} & \cellcolor{DarkOrange!40}\begin{tabular}[c]{@{}c@{}}AF31\\{[\car]}\end{tabular} & \multicolumn{2}{c|}{\cellcolor{LightGray!40}} \\ 
\hhline{-->{\arrayrulecolor{LightGray!40}}*{13}{-}>{\arrayrulecolor{black}}-->{\arrayrulecolor{LightGray!40}}-->{\arrayrulecolor{black}}}
\cellcolor{DarkSeaGreen!40}\textbf{FP10} & \multicolumn{1}{l|}{\cellcolor{DarkSeaGreen!40}\textbf{\begin{tabular}[l]{@{}l@{}}Lateral Distance\\Altering (A5)\end{tabular}}} & \multicolumn{2}{c|}{\cellcolor{LightGray!40}} & \multicolumn{4}{c|}{\cellcolor{LightGray!40}} & \multicolumn{3}{c|}{\cellcolor{LightGray!40}} & \multicolumn{2}{c|}{\cellcolor{LightGray!40}} & \multicolumn{2}{c|}{\cellcolor{LightGray!40}} & \multicolumn{1}{c|}{\textbf{\batk}} & \cellcolor{DarkOrange!40}\begin{tabular}[c]{@{}c@{}}AF32\\{[\car]}\end{tabular} & \multicolumn{2}{c|}{\cellcolor{LightGray!40}} \\ 
\hhline{-->{\arrayrulecolor{LightGray!40}}*{13}{-}>{\arrayrulecolor{black}}----}
\cellcolor{DarkSeaGreen!40}\textbf{\begin{tabular}[c]{@{}c@{}}FP11,\\12,13\end{tabular}} & \multicolumn{1}{l|}{\cellcolor{DarkSeaGreen!40}\textbf{\begin{tabular}[l]{@{}l@{}}Target Goal\\Generation (C2)\end{tabular}}} & \multicolumn{2}{c|}{\cellcolor{LightGray!40}} & \multicolumn{4}{c|}{\cellcolor{LightGray!40}} & \multicolumn{3}{c|}{\cellcolor{LightGray!40}} & \multicolumn{2}{c|}{\cellcolor{LightGray!40}} & \multicolumn{2}{c|}{\cellcolor{LightGray!40}} & \multicolumn{2}{c|}{\cellcolor{LightGray!40}} & \multicolumn{1}{c|}{\cite{spoof-voice-wc19,spoof-voice-inaudible-tdsc21}} & \cellcolor{DarkOrange!40}\begin{tabular}[c]{@{}c@{}}AF28-30\\{[\car]}\end{tabular} \\ 
\toprule[1.0pt]
\bottomrule[1.0pt]
\cellcolor{DarkSeaGreen!40} & \multicolumn{1}{l|}{\cellcolor{DarkSeaGreen!40}\textbf{Target Blurring (A3)}} & \multicolumn{2}{c|}{\cellcolor{LightGray!40}} & \multicolumn{1}{c|}{\textbf{\ratk}} & \multicolumn{1}{c|}{\textbf{\ratk}} & \multicolumn{1}{c|}{\textbf{\ratk}} & \cellcolor{DarkOrange!40} & \multicolumn{1}{c|}{\textbf{\ratk}} & \multicolumn{1}{c|}{\textbf{\ratk}} & \cellcolor{DarkOrange!40} & \multicolumn{1}{c|}{\cite{spoof-gps-car-iccv21}} & \cellcolor{DarkOrange!40} & \multicolumn{1}{c|}{\cite{spoof-imu-sp21}} & \cellcolor{DarkOrange!40} & \multicolumn{2}{c|}{\cellcolor{LightGray!40}} & \multicolumn{2}{c|}{\cellcolor{LightGray!40}}\\ 

\hhline{>{\arrayrulecolor{DarkSeaGreen!40}}->{\arrayrulecolor{black}}->{\arrayrulecolor{LightGray!40}}-->{\arrayrulecolor{black}}--->{\arrayrulecolor{DarkOrange!40}}->{\arrayrulecolor{black}}-->{\arrayrulecolor{DarkOrange!40}}->{\arrayrulecolor{black}}->{\arrayrulecolor{DarkOrange!40}}->{\arrayrulecolor{black}}->{\arrayrulecolor{DarkOrange!40}}->{\arrayrulecolor{LightGray!40}}---->{\arrayrulecolor{black}}}
\multirow{-2}{*}{\cellcolor{DarkSeaGreen!40}\textbf{FP14}} & \multicolumn{1}{l|}{\cellcolor{DarkSeaGreen!40}\textbf{ROI Altering (A3)}} & \multicolumn{2}{c|}{\cellcolor{LightGray!40}} & \multicolumn{1}{c|}{\textbf{\ratk}} & \multicolumn{1}{c|}{\textbf{\ratk}} & \multicolumn{1}{c|}{\textbf{\ratk}} & \multirow{-2}{*}{\cellcolor{DarkOrange!40}\begin{tabular}[c]{@{}c@{}}AF39-41\\{[\mb\,\drone]}\end{tabular}} & \multicolumn{1}{c|}{\textbf{\ratk}} & \multicolumn{1}{c|}{\textbf{\ratk}} & \multirow{-2}{*}{\cellcolor{DarkOrange!40}\begin{tabular}[c]{@{}c@{}}AF36-38\\{[\mb\,\drone]}\end{tabular}} & \multicolumn{1}{c|}{\cite{spoof-gps-car-autosec21}} & \multirow{-2}{*}{\cellcolor{DarkOrange!40}\begin{tabular}[c]{@{}c@{}}AF42-44\\{[\car\,\drone]}\end{tabular}} & \multicolumn{1}{c|}{\textbf{\batk}} & \multirow{-2}{*}{\cellcolor{DarkOrange!40}\begin{tabular}[c]{@{}c@{}}AF33-35\\{[\mb\,\drone]}\end{tabular}} & \multicolumn{2}{c|}{\cellcolor{LightGray!40}} & \multicolumn{2}{c|}{\cellcolor{LightGray!40}} \\
\hline
\end{tabular}}
\begin{adjustwidth}{0.1cm}{0.1cm}
\begin{tablenotes}[para,flushleft]
\footnotesize
    \centering \textbf{\batk/\gatk/\ratk}:\;Unexplored attacks \quad \textbf{LP/SM/OP}:\;Laser Projection/Shape Manipulation/Object Placement \quad \textbf{SP/LP}:\;Sticker Pasting/Light Projection \\ 
    \revision{\textbf{A1}: Object detection \enspace \textbf{A2}: Segmentation \enspace \textbf{A3}: Localization/SLAM \enspace \textbf{A4}: Speech recognition \enspace \textbf{A5}: Distance detection \enspace \textbf{B}: Object tracking} \\
    \revision{\textbf{C1}: Environment prediction \enspace \textbf{C2}: Goal planning \enspace \textbf{D1}: Path routing \enspace \textbf{D2}: Parking/Landing planning \enspace \textbf{E}: Motion planning \enspace \textbf{F}: Motion controller} \\
    \revision{\textbf{FP}:\;Flow Pattern \quad \textbf{AF}:\;Action Flow}
\end{tablenotes}
\end{adjustwidth}
\label{tb:3-fp}
\vspace{-15pt}
\end{table*}

The five functions (\textit{Objection Detection} -- \textit{Distance Detection}) are the main targets for the adversary to compromise the RV system. We define the action flow \revision{(ActFlow)} as a complete flow from one sensor to a final control function\footnote{An action flow \revision{(ActFlow)} provides more information than an attack path in \cref{sec:2-AP} since it is at the granularity of robotic functions}. Then we hypothesize that \textit{each action flow could give rise to some sensor spoofing vectors}, where the adversary can tamper with the corresponding sensor data to affect RV's final actions. Table \ref{table:action_flow} in the appendix lists 44 possible action flows with the corresponding attack paths. A single-round action flow refers to the flow going from the perception to control stages once, while a multi-round action flow goes through the pipeline multiple times with the interaction between the RV and environment. We observe that some action flows have been exploited to launch sensor spoofing attacks, while most have not been investigated yet, leaving a large unknown attack surface for RV systems. 

We obtain three interesting observations from the summary of these action flows and attack vectors. (1) Most existing works design the attacks heuristically, and they mainly focus on the immediate action damage from the mis-estimation of RV states. (2) Some action flows require specific function compositions and sensor types. They can only be launched against specific types of RVs. (3) Spoofing attacks based on multi-round action flows are rarely considered by prior works due to their complex mechanisms. Only three papers discussed these attacks. Below we give detailed analysis about each type of spoofing attacks. 

\subsection{Attack Vector Analysis}
\label{sec:3-atk}
Some action flows may share the same composition and interactions of robotic functions, and only differ in the target sensors. The corresponding spoofing attacks will have common characteristics. To simplify the analysis, we treat these action flows as the same \textit{flow pattern \revision{(FlowPtn)}}. Then the 44 action flows lead to 14 different flow patterns.

\begin{figure*}[tb]
\centering
\includegraphics[width=1.00\linewidth]{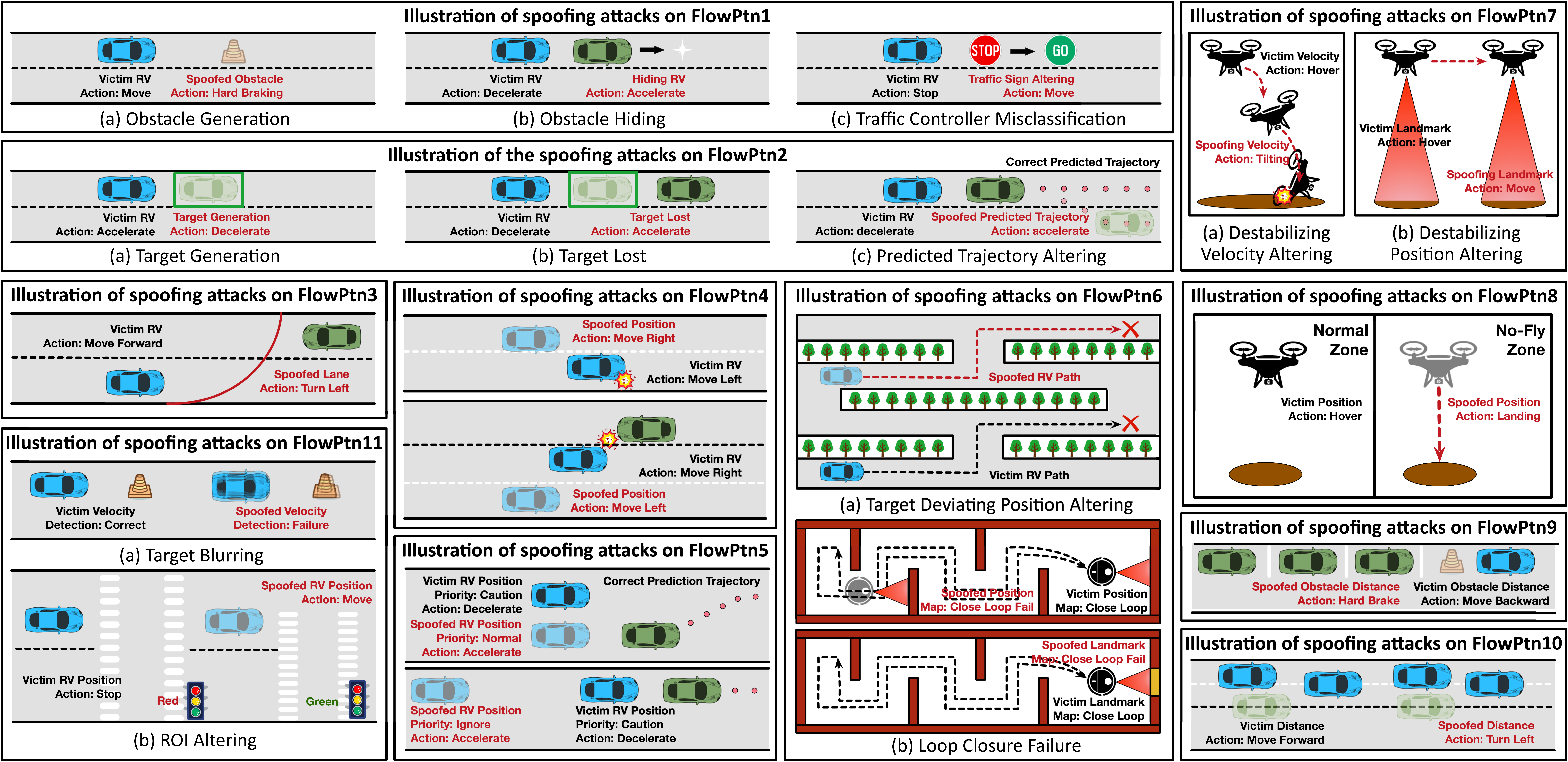}
\vspace{-15pt}
\caption{Illustration of spoofing attacks on flow pattern 1-11.}
\vspace{-20pt}
\label{fig:apx-atk-figures}
\end{figure*}

Table \ref{tb:3-fp} summarizes these attacks with different target sensors and spoofing techniques. Each attack falls into one of two cases: 
(1) \revisionB{there are already existing works realizing this attack. We list the references in the corresponding cells. 
Most of these works have been empirically verified and tested on simulators (e.g., Baidu Apollo \cite{rv-system-apollo}) or physical world.} 
\xuyuan{
(2) There is no existing work exploring this attack, yet it is possible and realistic. In \cref{sec:3-analysis}, we further analyze and categorize them into three cases based on the attack feasibility (\batk/\gatk/\ratk). 
}
Below we discuss the mechanism of each flow pattern, while detailed attack scenarios are illustrated in Figure \ref{fig:apx-atk-figures}. 

\noindent\textbf{1) 
\revision{FlowPtn 1}} \hyperref[fig:3-workload-model]{(\emph{\revision{Object Detection}$\rightarrow$\revision{Motion Planning}$\rightarrow$ \revision{Motion Controller}}):}
This flow pattern contains the action flows of \revision{ActFlow1,3,10}. They are responsible for avoiding obstacles and taking correct actions based on \revisionC{the} traffic controller, such as traffic lights and signs. To achieve correct motion planning, the RV needs to obtain the accurate position and property of these traffic controllers. We summarize three possible attacks that can manipulate the objects and compromise the executions
(Figure \ref{fig:apx-atk-figures}.\revision{FlowPtn1}). 
(i) \emph{Obstacle Appearing}: the RV mistakenly recognizes a non-existent obstacle in front. Then it brakes hard and \revisionC{stops} at the intersection even if the traffic light is green. (ii) \emph{Obstacle Missing}: the RV fails to detect an existing obstacle in front, and crash into it directly. 
(iii) \xingshuo{\emph{Traffic Controller Misclassification}: the RV misclassifies the traffic sign or traffic light into a wrong category and takes dangerous actions.}

According to Table \ref{tb:3-fp}, we observe that a number of attack vectors have been realized in prior works while some have never been explored. Specifically, there is no related work focusing on obstacle appearing attack using shape manipulation, object placement and sticker pasting spoofers. Moreover, for the obstacle missing, one possible attack is to simply use radar absorbing material to implement. Recently, \revisionC{LiDAR} is also used for traffic controller classification \cite{weng2016road}, but no work explores misclassification attack using \revisionC{LiDAR} spoofers.



\noindent\textbf{2) 
\revision{FlowPtn 2}} \hyperref[fig:3-workload-model]{(\emph{\revision{Object Detection}$\rightarrow$\revision{Object Tracking}$\rightarrow$\revision{Env Prediction}$\rightarrow$\revision{Motion Planing}$\rightarrow$\revision{Motion Controller}}):}
This pattern includes action flows \revision{ActFlow2,4,11}. They focus on tracking the dynamic targets and predicting their trajectories. Similar to Flow Pattern 1, we consider three possible attacks that manipulate the tracked target or predicted trajectory (Figure \ref{fig:apx-atk-figures}.\revision{FlowPtn2}).
(i) \emph{Trajectory Appearing}: the adversary can fool the RV to assign a track id to a non-existent object and then brake hard to avoid hitting it. 
(ii) \emph{Trajectory Missing}: the spoofer can make the victim RV loss the tracking target, which could shorten the safe distance and cause the RV to crash into the target. 
(iii) \emph{Trajectory Altering}: the adversary alters the predicted trajectory of the target. This can also reduce the safe distance with potential vehicle crashes. 

From Table \ref{tb:3-fp}, we can observe that recent work \cite{spoof-radar-tifs21} has discussed the possibility of trajectory appearing using the radar spoofer and only one related work \cite{spoof-camera-stick-iclr20} has designed all these three attacks through the sticker pasting spoofer. However, the implementation using other sensor spoofers are never considered. 


\noindent\textbf{3)  
\revision{FlowPtn 3}} \hyperref[fig:3-workload-model]{(\emph{\revision{Segmentation}$\rightarrow$\revision{Env Prediction}$\rightarrow$\revision{Motion Planning}$\rightarrow$\revision{Motion Controller}}):}
This flow pattern includes \revision{ActFlow12}, which identifies the road conditions (e.g., lanes) and  guides RV's motion control. Thus, one possible attack is \emph{Lane Altering}, which makes the RV identify wrong traffic lanes (Figure \ref{fig:apx-atk-figures}.\revision{FlowPtn3}). 
Past works have demonstrated such attack using sticker pasting \cite{spoof-camera-stick-us21,spoof-camera-stick-us21-cq} and light projection spoofing \cite{spoof-camera-light-iacr20}. 

\noindent\textbf{4) 
\revision{FlowPtn 4}} \hyperref[fig:3-workload-model]{(\emph{\revision{Localization/SLAM}$\rightarrow$\revision{\mbox{Motion Planning}} $\rightarrow$\revision{Motion Controller}}):}
This flow pattern (\revision{ActFlow8,16,21,26}) assists the localization of the RV. For example, autonomous vehicles need to drive in the center of the lane for safety. The \textit{Localization} function helps the vehicle determine if it is on the right track. Thus, one possible attack is \emph{Deviating Position Altering}, which exploits this lateral deviation to cause mis-prediction of RV's location. 
As shown in Figure \ref{fig:apx-atk-figures}.\revision{FlowPtn4}, the RV is spoofed to a position away from the lane. Due to the false position, 
the RV takes actions to move in the opposite direction and further drive off the road pavement or on the wrong ways. Table \ref{tb:3-fp} summarizes that such attack has only been realized by the GPS spoofer \cite{spoof-gps-car-us20}. 
\xuyuan{
Due to the different demands on the cost and scenario, some \textit{Localization} functions only depend on LiDAR \cite{rv-lidar-loc}, camera \cite{rv-camera-loc} or IMU sensor \cite{rv-imu-loc}, thus spoofing on these sensors can also achieve the same attack consequence. 
}

\noindent\textbf{5) 
\revision{FlowPtn 5}} \hyperref[fig:3-workload-model]{(\emph{\revision{Localization/SLAM}$\rightarrow$\revision{Env Prediction}$\rightarrow$ \revision{Motion Planning}$\rightarrow$\revision{Motion Controller}}):}
This pattern covers the action flows of \revision{ActFlow5,13,18,23}. They predict whether the RV should take actions due to surrounding objects. The predicted outcome is one of three priorities: \emph{ignore}, \emph{caution} and \emph{normal}. The first two indicate the object will not and will most likely affect RV's trajectory, while the last one indicates other conditions by default. Thus, one possible attack is \emph{Predicted Priority Altering}, which misleads the RV from the caution priority to the ignore priority, thus taking dangerous actions subsequently (Figure \ref{fig:apx-atk-figures}.\revision{FlowPtn5}). It can be launched by targeting LiDAR, camera, GPS or IMU sensors. There are no prior works realizing such attack vector.

\noindent\textbf{6) 
\revision{FlowPtn 6}} \hyperref[fig:3-workload-model]{(\emph{\revision{Localization/SLAM}$\rightarrow$\revision{Path Routing}$\rightarrow$ \revision{Motion Planning}$\rightarrow$\revision{Motion Controller}}):}
These action flows (\revision{ActFlow6,14,19,24}) control the RV to navigate from one location to another or map an unknown area, i.e., SLAM. Two possible attacks are introduced by generating wrong paths or causing failure of the mapping task: (i) \emph{Target Deviating Position Altering}: the RV is guided to a wrong destination (Figure \ref{fig:apx-atk-figures}.\revision{FlowPtn6}(a));
(ii) \emph{Loop Closure Failure}: the RV fails to assert that it returns to a previously visited location so that the map cannot be correctly generated (Figure \ref{fig:apx-atk-figures}.\revision{FlowPtn6}(b)). 

From Table \ref{tb:3-fp}, the target deviation position altering attack has been realized in \cite{spoof-gps-car-us18}, which adopts the GPS spoofer to slightly shift RV's position to make the fake navigation route match the shape of the actual roads. 
\xuyuan{
As we discussed in FlowPtn 4, spoofing other sensors can achieve the same consequence as well, especially for indoor RVs without GPS sensors. 
}
There are no prior works realizing the loop closure failure attack. We can use the GPS spoofer to mislead the RV to a fake position when recognizing the previous map, or use other sensor spoofers to directly modify the perceived object to mismatch the previous landmarks (\cref{sec:4-case-2}).

\noindent\textbf{7) 
\revision{FlowPtn 7}} \hyperref[fig:3-workload-model]{(\emph{\revision{Localization/SLAM}$\rightarrow$\revision{Motion Controller}}):}
These action flows (\revision{ActFlow9,17,22,27}) aim to stabilize the RV based on the variance of the velocity or position. For example, drones need to dynamically adjust the fuselage to prevent overturning, or adjust its position to maintain hovering according to the changes of the ground below. 
Thus, two possible attacks can be proposed to destabilize the RV (Figure \ref{fig:apx-atk-figures}.\revision{FlowPtn7}).
(i) \emph{Destabilizing Velocity Altering}: the adversary forges an angular velocity in a single direction and makes the RV overturned. 
(ii) \emph{Destabilizing Position Altering}: the spoofer creates a continuous slight difference on the ground that misleads the RV to laterally move to a designated position. 

As presented in Table \ref{tb:3-fp}, many works use the IMU spoofer to alter the destabilizing velocity \cite{spoof-imu-eurosp17,spoof-imu-blackhat17,spoof-imu-asiaccs18,spoof-imu-us18}. 
\xuyuan{
Since the camera \cite{rv-camera-odom} or LiDAR \cite{rv-lidar-odom} is also used to estimate odometry through matching the position of extracted features in two adjacent frames/point cloud data, spoofing these sensors is another way to achieve such attacks. To the best of our knowledge, GPS is rarely used as the odometry estimator. 
}
Davidson et al. \cite{spoof-camera-light-woot16} realized the destabilizing position altering attack on a drone using a light projection spoofer. 
\xuyuan{
If an adversary can slightly move a sticker pasting spoofer on the ground while the drones hovers, such attack can also be achieved. Besides, spoofing false \revisionC{LiDAR} points or GPS signals can modify the hovering height.
}

\noindent\textbf{8) 
\revision{FlowPtn 8}} \hyperref[fig:3-workload-model]{(\emph{\revision{Localization/SLAM}$\rightarrow$\revision{Parking/Landing Planning}$\rightarrow$\revision{Motion Planning}$\rightarrow$\revision{Motion Controller}}):}
This pattern covers the action flows of \revision{ActFlow7,15,20,25}, which control specific RV tasks, such as parking or landing. To complete the action flow, the RV is provided with one target location. Thus, one possible attack is \emph{Specific Location Altering}, where the adversary triggers the launch of these tasks at wrong time or locations. As shown in Figure \ref{fig:apx-atk-figures}.\revision{FlowPtn8}, by spoofing a location to a no-fly zone, the victim drone is enforced to perform an immediate landing. Table \ref{tb:3-fp} shows related works on such attack based on the GPS spoofer. 
\xuyuan{
If an adversary needs to implement such attack on the RV using LiDAR or camera for localization, he can construct a similar environment to cheat the loop-closure system by the LiDAR or camera spoofer. Besides, spoofing IMU to accumulate position errors can also achieve the same effect.
}

\noindent\textbf{9) 
\revision{FlowPtn 9}} \hyperref[fig:3-workload-model]{(\emph{\revision{Distance Detection}$\rightarrow$\revision{Parking/Landing Planning}$\rightarrow$\revision{Motion Planning}$\rightarrow$\revision{Motion Controller}}):}
This flow pattern refers to \revision{ActFlow31}, which ensures the safe distance between the RV and obstacles in the parking or landing tasks. Thus, one possible attack is \emph{Obstacle Distance Altering}, which uses the ultrasonic sensor spoofer to shorten this distance and cause the RV to brake hard (Figure \ref{fig:apx-atk-figures}.\revision{FlowPtn9}) \cite{spoof-ultrosonic-defcon16,spoof-ultrosonic-iotj18}.

\noindent\textbf{10) 
\revision{FlowPtn 10}} \hyperref[fig:3-workload-model]{(\emph{\revision{Distance Detection}$\rightarrow$\revision{Motion Controller}}):}
The action flow \revision{ActFlow32} is responsible for ensuring a lateral safe distance between the RV and nearby cars encroaching on its lane. Therefore, we propose a new attack: \emph{Lateral Distance Altering}. As shown in Figure \ref{fig:apx-atk-figures}.\revision{FlowPtn10}, by deploying many ultrasonic spoofers along the roadside, the vehicle needs to frequently change different directions to ensure it is safe within the designated road lane. This could make the autonomous driving less smooth, and annoy or even hurt the passengers. This attack is realizable but no prior works ever considered it (Table \ref{tb:3-fp}).

\noindent\textbf{11) 
\revision{FlowPtn 11,12 and 13}} \hyperref[fig:3-workload-model]{(\emph{\revision{Speech Recognition}$\rightarrow$\revision{Goal Planning}$\rightarrow$[\revision{Path Routing}/\revision{Parking/Landing Planning}: \revision{Motion Planning}]/-$\rightarrow$\revision{Motion Controller}}):}
These three action flows focus on launching and performing various missions according to the user's voice commands. They are vulnerable to the \emph{Target Goal Generation} attack, which triggers malicious missions using a microphone spoofer. It has been implemented in prior works \cite{spoof-voice-wc19,spoof-voice-inaudible-tdsc21}.

\noindent\textbf{12) 
\revision{FlowPtn 14}} \hyperref[fig:3-workload-model]{(\emph{\revision{Localization/SLAM}$\rightarrow$\revision{Motion Controller} \ding{242} \revision{Object Detection}}):}
Finally, we consider multi-round action flows\footnote{In this paper, we only consider two-round \revision{ActFlows}. Attacks with more rounds are more complex, and will be investigated as future work.}. There are two observations that lead to two attack vectors. (i) The quality of the perceived images or laser points highly depends on the stabilization of the RV. For example, the jitters of camera or LiDAR can cause blurred images and irregular distributions of point clouds. Hence, we have the \emph{Target Blurring} attack, which spoofs the sensors to jitter the RV and cause the failure of object detection in LiDAR or camera (Figure \ref{fig:apx-atk-figures}.\revision{FlowPtn11}(a)). (ii) The ROI depends on the current position of the RV. Therefore, we have the \emph{ROI Altering} attack, which falsifies well-designed positions and alters the detected ROI. As shown in Figure \ref{fig:apx-atk-figures}.\revision{FlowPtn11}(b), the vehicle mis-estimates the traffic signals based on the second traffic light rather than the first one due to the spoofed position.

As shown in Table \ref{tb:3-fp}, past works realize the target blurring attacks against the motion compensation mechanism in autonomous vehicles to blur the images \cite{spoof-imu-sp21} and point clouds \cite{spoof-gps-car-iccv21}. For other RVs that do not adopt this compensation mechanism (e.g., indoor automated guided vehicles and drones), we can use other spoofers to cause jitters. Tang et al. \cite{spoof-gps-car-autosec21} implemented an ROI altering attack with the GPS spoofer. 
\xuyuan{
Also, spoofing other location-related sensors can also achieve the same consequence.
}

\subsection{Feasibility Analysis}
\label{sec:3-analysis}


\xuyuan{
\revisionC{With the action flow model, we have identified} \revisionS{77 unexplored threats} \revisionC{that are not considered in prior works and have the potential to cause fatal accidents}. Below we discuss the feasibility of these attacks from three cases. 
}

\vspace{3pt}
\noindent
\textbf{C1. Learn from each other} (\gatk): 
In LiDAR spoofing attacks, we summarize three techniques: laser projection, shape manipulation and object placement. They are actually able to achieve the same compromised state (i.e., spoofing the same points) by using an active LiDAR spoofer, creating and placing an adversarial object, respectively. As a result, a new attack based on one spoofing technique is also feasible if this attack has been realized with another technique. For example, in \revision{FlowPtn1}, the implementation of obstacle appearing attack using LiDAR object placement can be inspired by the same attack with laser projection \cite{spoof-lidar-bhe15,spoof-lidar-shin,jin2022pla,spoof-lidar-ccs19,spoof-lidar-us20}. Similarly, the two techniques for camera sensors (sticker pasting and light projection) can also target the same perception function, and their attacks have similar feasibility. 

\vspace{3pt}
\noindent
\textbf{C2. Easy to prove} (\batk): 
\xingshuo{(1) The obstacle missing attack (\revision{ActFlow1}) and trajectory missing attack (\revision{ActFlow2}) using radar can be simply implemented by using radar absorbing materials. Moreover, trajectory altering attack can also be implemented by combining existed appearing attacks.
(2) Traffic controller misclassification attack using  LiDAR (\revision{ActFlow3}) can be easily transferred from camera. For example, an adversary can leverage a laser beam to change the semantic information of a traffic sign. }
\xuyuan{
(3) Loop closure failure attack (\revision{FlowPtn6.SLAM}). The SLAM algorithm helps the RV recognize a visited location. By adding enough visual noise in the critical region of the environment, the RV will fail to detect the loop closure and then generate the wrong map. 
(4) Destabilizing velocity altering attacks based on LiDAR or camera spoofing (\revision{ActFlow9,17}). The LiDAR- or camera-based odometer estimates RV's velocity based on the offset of the detected point clouds/pixels. As prior works \cite{spoof-lidar-camera-sp21,spoof-lidar-ccs21,spoof-camera-stick-ccs19,cheng2022physical,huang2020universal,wu2020making,xu2020adversarial,spoof-camera-light-us21} have shown that both two sensors can be spoofed to lose the target (\revision{FlowPtn1}), these techniques can also cause wrong estimates of the odometer to realize these attacks. 
(5) Attacks based on GPS spoofing (\revision{ActFlow18,19,22}). GPS spoofing is more mature than other spoofing attacks, thus predicted priority altering attack can be easily achieved (\revision{ActFlow18}). Note that GPS is rarely used as the main sensor for mapping (\revision{ActFlow19}) and odometry (\revision{ActFlow22}).
(6) Attacks based on IMU spoofing (\revision{ActFlow23,24,26,33-35}). These attacks aim to alter RV's position. They can be realized following the destabilizing velocity altering attack using the IMU spoofer (\revision{ActFlow27}), \mbox{which can cause the accumulative error. }
}

\vspace{3pt}
\noindent
\textbf{C3. Need to validate} (\ratk): 
(1) Trajectory appearing, missing and altering attacks based on LiDAR spoofing (\revision{FlowPtn2}, \revision{ActFlow4}). Past work \cite{spoof-camera-stick-iclr20} only demonstrated a camera-based sticker pasting method for these attacks (\revision{ActFlow11}) . It is unknown whether we can spoof the LiDAR to achieve the same result.  
(2) Attacks based on camera spoofing (\revision{ActFlow13,14,16,36-38}) or LiDAR spoofing (\revision{ActFlow5,6,8,39-41}). They are built on the assumption that the SLAM function can be fooled to re-localize the RV to a false position. To show their practicality, we design two novel methodologies and implement prototypes in the next section. 
\section{Two Novel Attack Methodologies}
\label{sec:4}
We present two new approaches to validate the feasibility of attack vectors in case 3 (\cref{sec:3-analysis}). We believe each approach has sufficient technical novelty and contributions as an individual research project. Due to the page limit, we only describe the basic mechanism and evaluation. We expect researchers can extend them to design more spoofing attacks, since they are the basis of the 36 unexplored vectors in case 3. 

\subsection{Obstacle Position Altering (LiDAR)}
\label{sec:4-case-1}
RVs utilize state-of-the-art DNN models to interpret 3D point clouds captured by the LiDAR for object detection. 
In general, a 3D object detection model $\mathcal{M}$ extracts features from the input point cloud $X$, and outputs a set of bounding boxes $Y = \mathcal{M}(X)$. Each box is attached to a detected object with its location $Y_{loc}$, size $Y_{size}$, heading $Y_{h}$, and confidence score $C$ of predicted categories. Boxes with the confidence score lower than a threshold will be filtered out, and the remaining box is the detected object. 

\begin{figure}[tb]
\centering
\includegraphics[width=1.0\linewidth]{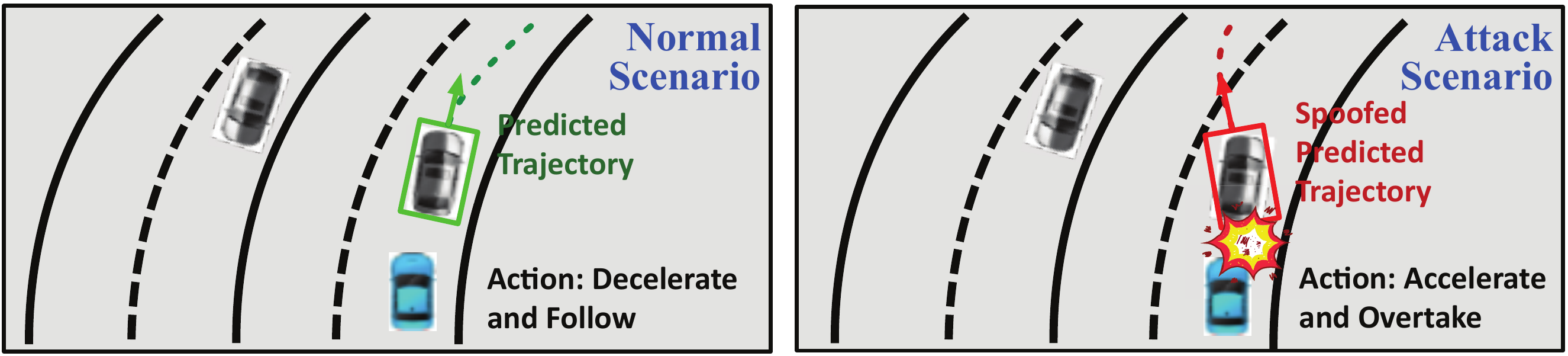}
\vspace{-15pt}
\caption{An example of obstacle position altering.}
\vspace{-15pt}
\label{fig:case_1_1}
\end{figure}

The goal of our attack is to fool $\mathcal{M}$ to mis-recognize a moving object in a wrong location, and then the RV will mis-estimate the object's trajectory. Figure \ref{fig:case_1_1} shows an example. There is a running vehicle in front of the victim RV. The adversary can spoof the point clouds to mislead the victim that this vehicle was switching to the left lane. Then the RV will accelerate, and the safety distance will be reduced, which can cause car crash (\revision{FlowPtn2,4}). Note this attack is different from existing LiDAR spoofing attacks \cite{spoof-lidar-bhe15,spoof-lidar-shin,jin2022pla,spoof-lidar-ccs19,spoof-lidar-us20,spoof-lidar-camera-sp21,spoof-lidar-ccs21}, which manipulate the existence of static obstacles. Our attack aims to alter the predicted trajectory of a moving object. Although the same attack goal has been realized by camera spoofing \cite{spoof-camera-stick-iclr20}, our attack targets the LiDAR-based RVs (e.g., Baidu Apollo), and is technically more challenging due to the complex and non-differentiable feature of point cloud models.


Formally, we inject an adversarial object $x^*$ into the original input $X$. The model output $Y^* = \mathcal{M}(X+x^*)$ has the same size and category as $Y$. However, its location $Y^*_{loc}$ and heading $Y^*_{h}$ are different. To craft a qualified $x^*$, we adopt a common object (e.g., drone) and try to find a malicious placement $s^*=(x^*,y^*,z^*,\alpha^*, \beta^*, \gamma^*)$ in the environment, where the first three variables denote its location, and the last three denote its orientation. $s^*$ can maximize the distance between the predicted and original locations of the bounding box:
\vspace{-2pt}
\begin{equation}
    \max_{s^*} ||Y^*_{loc}, Y_{loc}||
\end{equation}
\vspace{-2pt}
We can use an optimization method to identify $s^*$. The challenge is that the optimization objective is non-differentiable, and it is hard to calculate the gradient. To address this issue, we follow~\cite{kariyappa2021maze} to estimate the gradient as below: 
\vspace{-2pt}
\begin{equation}
\label{equ:altering}
\hat{\bigtriangledown} _{s^*}{\mathit{L}_{\mathcal{M}}}(s^*)= \frac{1}{m}\sum_{i=1}^{m}\frac{{\mathit{L}_{\mathcal{M}}}(s^*+\epsilon u_{i})*u_{i}}{\epsilon }
\end{equation}
\vspace{-2pt}
where $u_i$ is a random variable with a uniform distribution; $\epsilon$ is a positive smoothing factor; $\mathit{L}_{\mathcal{M}}$ is the loss function to quantify the MSE distance between $Y^*_{loc}$ and $Y_{loc}$. \revisionC{$m$ is a hyperparameter to control the gradient estimation.} 

Algorithm~\ref{altering} details the optimization process. The adversary iteratively adjusts the location and orientation of the adversarial object to affect the bounding box of the target obstacle. In each iteration, we estimate the gradient with  Eq.~\ref{equ:altering}, and use Project Gradient Descent (PGD) to update the gradient.

\begin{algorithm}
\caption{Position Altering Attack}
\footnotesize
\label{altering}
\begin{algorithmic}[1]
\Require clean point cloud $X$; \# of attack iterations $N$; loss threshold $\theta$;
target object to be altered $T$; adversarial object $x^*$
\Ensure adversarial object placement $s^*$.
\State Initialize: $s^*$
\For{$j=0$; $j<N$; $j++$}
    \State $grad = 0; counter = 0; i = 0$
    \While {$i < m$}
        \State $loss = \mathit{L}_{\mathcal{M}}(s^* + \epsilon * u_i)$
        \If{$loss > \theta$}
            \State $i = i + 1$
            \State $grad = grad + (loss / \epsilon) * u_i$
        \EndIf
        \State $grad = grad / m$
        \State $s^* += sign(grad) * \epsilon$
    \EndWhile
\EndFor
\end{algorithmic}
\end{algorithm}

\begin{figure}[tb]
\centering
\includegraphics[width=1.0\linewidth]{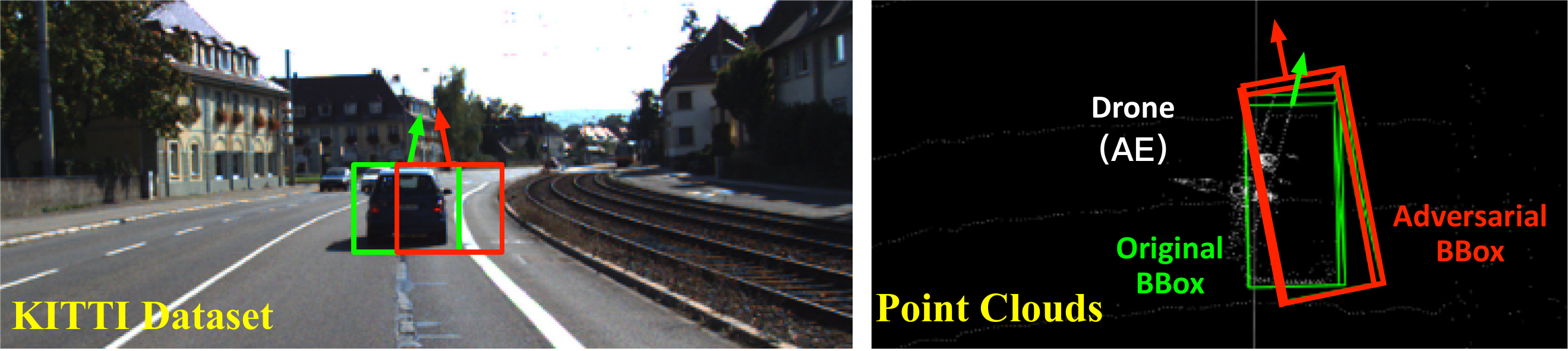}
\vspace{-15pt}
\caption{Position altering attack on PointRCNN.}
\vspace{-15pt}
\label{fig:case_1_2}
\end{figure}

It is worth noting that we focus on the trajectory altering attack in \revision{ActFlow4}. There are also two attacks (trajectory appearing and missing) in this action flow. They can be achieved using the same method by changing the optimization objective adaptively: instead of maximizing the location distance, we can manipulate the confidence score $C$ of the bounding boxes to create a non-existence target, or hide an existing target. Due to the page limit, we leave this as future work. 

\noindent
\textbf{Experiments}. We choose PointRCNN~\cite{shi2019pointrcnn}, one of the most widely used LiDAR detection models. We use the popular KITTI dataset \cite{kitti} to train this model, where the samples are collected by the Velodyne HDL-64E Lidar. We use a drone as the adversarial object due to its flexibility and stealthiness~\cite{spoof-lidar-ccs21}. The location of the drone is restricted to a space of 3m * 3m * 1m around the target vehicle. We set $\epsilon = 0.1$, and $s^*$ is initialized as random values between $[-1, 1]$.

\revision{Figure~\ref{fig:case_1_2} shows the outcome of our position altering attack. The objective of this attack was to deceive the victim vehicle's LiDAR and cause a misestimation of the position of the front vehicle. This was achieved by utilizing a drone and iteratively optimizing its pose through 50 iterations using Algorithm~\ref{altering}. We can observe the PointRCNN model in the victim vehicle was successfully manipulated, leading to a wrong prediction of the position of the front vehicle shifting to the left. This incorrect estimation could result in the victim vehicle accelerating and overtaking, potentially causing a rear-end collision.} 


\subsection{Location Altering (Camera/LiDAR)}
\label{sec:4-case-2}
SLAM is widely adopted in many RVs for localization and mapping. State-of-the-art SLAM algorithms take camera images or LiDAR point clouds as input. One key module in SLAM is loop closure, which checks if the RV has returned to a previously mapped region to reduce the uncertainty in the map estimation. Different algorithms share the similar idea in loop closure detection: the RV constructs the surroundings of the current location with the sensor data, and compares them with the established global map. A loop closure is detected when the similarity between the current environment and a part of the map is higher than a threshold. To attack the loop closure function, the adversary can modify the environment to increase the similarity between two distinct locations. Then the RV will mistakenly think a new location is visited before, and mis-estimate its location. This leads to a series of spoofing attacks (\revision{ActFlow5,6,8,13,14,16,36-41}).

\begin{figure}[tb]
\centering
\includegraphics[width=1.0\linewidth]{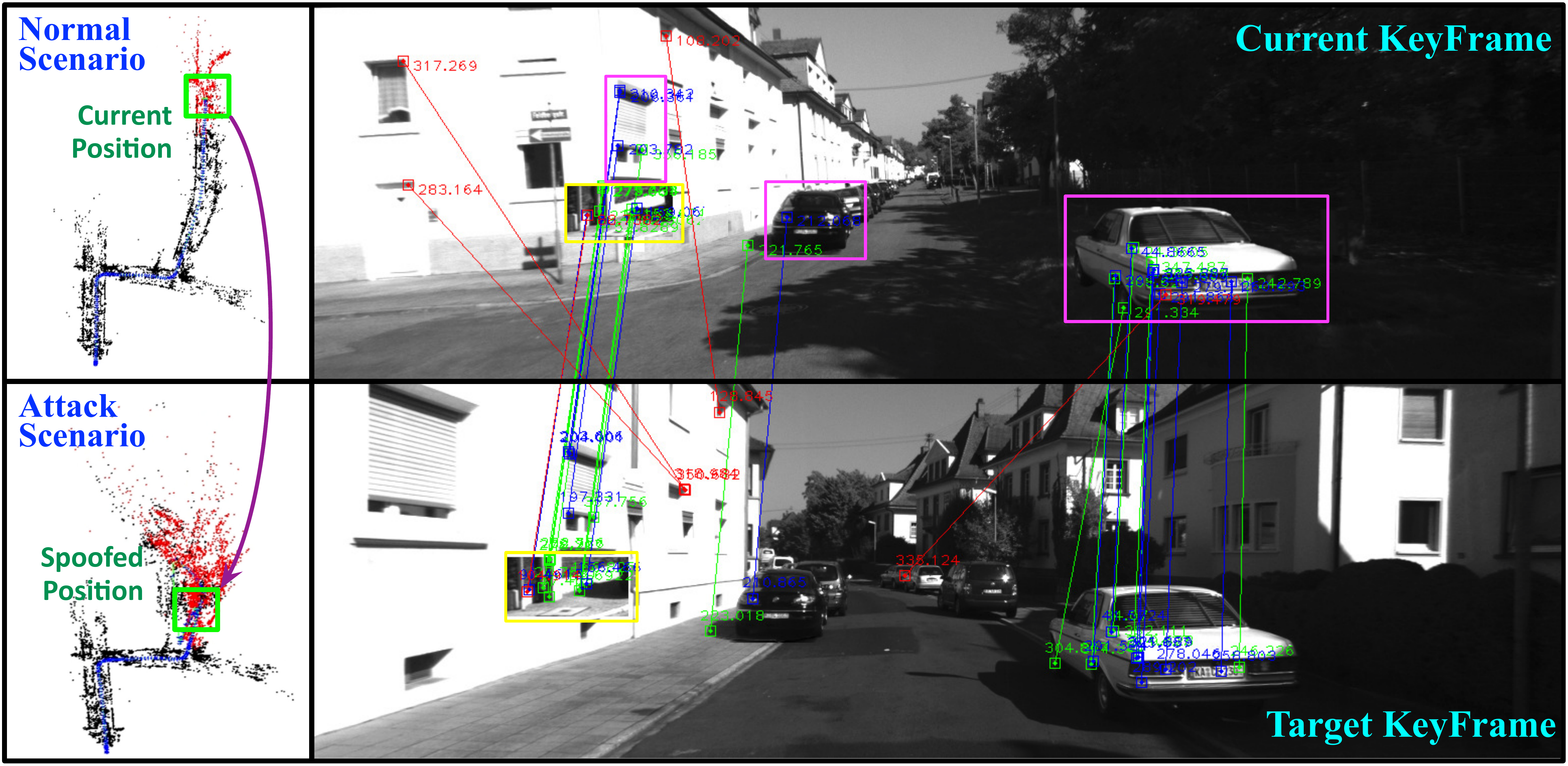}
\vspace{-15pt}
\caption{Failure of loop closure in the ORB-SLAM2 simulator.}
\vspace{-15pt}
\label{fig:4-slam-atk}
\end{figure}

\noindent
\textbf{Experiments}. 
We target ORB-SLAM2 \cite{orb-slam2}, the most popular real-time SLAM algorithm in both academia and industry. We implement this algorithm in a simulated environment with its default setting, and choose a mid-size city in KIITI \cite{kitti} as our experimental scenario. During the SLAM process, the vehicle extracts the FAST keypoints \cite{fast} and ORB descriptors \cite{orb} from each frame. It calculates the similarities between the current and past frames based on their relative scale, position and angles of the keypoints. A loop closure is reported if this similarity exceeds a pre-defined threshold.

As shown in Figure \ref{fig:4-slam-atk}, we try to cause false loop closure on two distinct locations: the current keyframe is a T-turn intersection, while the target keyframe is a straight road. We increase their similarity by two means: (1) moving certain physical objects from the target keyframe to the current keyframe (purple box); (2) adding some patches to both keyframes (yellow box). We hope to bring minimum changes to the environment, so we carefully select the objects and patches that are relatively small but contain many FAST keypoints. They are either printed out as patches to stick on the wall, or used as references to find the same objects and place them in the current frame. We maintain the relative scale and position of the selected objects  in two scenes so that they can effectively contribute to the similarity calculation.

After adding the patches and objects, we get 34 pairs of matched keypoints between the current and target keyframes (divided into 3 groups based on the relative angles, connected by green, blue and red lines, respectively), which fulfill the ORB-SLAM2 loop closure requirements. The green boxes in the left side of Figure \ref{fig:4-slam-atk} denote the current positions of the autonomous vehicle in the map. When the vehicle reaches the intersection, it mistakenly detects a loop closure, and relocates itself to the  position of the target frame. This verifies the feasibility of the location altering attack. 
\section{Discussion and Future Work}
\subsection{Attacks against Multi-Sensor Fusion}
This paper mainly discusses spoofing attacks with single sensors. Modern RV systems start to integrate Multi-sensor Fusion (MSF) algorithms to smooth out errors and uncertainties of each single sensor and improve the perception robustness. As our action flow model is built upon the interaction among robotic functions, it can describe these sophisticated attacks against MSF as well. 
For instance, 
some MSF-based object detection functions use LiDAR data to assist the camera to discern the depth range~\cite{rv-system-apollo,wang2019frustum,ku2018joint,huang2020epnet}. The adversary can use a single LiDAR spoofer to make the obstacle closer to the victim and cause collisions~\cite{hallyburton2021security}. Cao et al.\cite{spoof-lidar-camera-sp21} observed that the shape of a 3D object can cause position changes in LiDAR point clouds as well as pixel value changes in camera images. Then they proposed a spoofing method to blind the MSF-based object detection by optimizing a 3D-printed obstacles. Both two attacks can be categorized as action flow AF3. 

\subsection{Spoofing Defenses}
We focus on spoofing attacks, and the systematization of defense solutions is not covered in this paper. Generally, defense works can be classified into two categories: (1) Detection: the defender aims to identify the existence of the spoofing activities \cite{spoof-magnetic-ccs15} or fake signals \cite{bolton2018blue}. (2) Prevention: the defender tries to correct the spoofed data or state by filtering \cite{kune2013ghost}, randomization \cite{spoof-lidar-shin}, fusion \cite{spoof-ultrosonic-iotj18}, etc. Interested readers can refer to \cite{sok-analog-sp21} for more details. 

An interesting direction is to leverage our action flow model to build a unified defense. Since the model depicts the interactions between RV functions, it can also identify the key to mitigate corresponding attacks. Each flow pattern consists of multiple functions, and each function may be subject to a spoofing attack. So we can design methodologies for system monitoring at the function level, and combine them to detect anomalies in different flows. 
\revision{Our proposed defensive scheme focuses on using the interaction between different action flows to detect attacks, while the scheme in \cite{sok-analog-sp21} relies on the feature of each victim function for monitoring and prevention, such as adaptive filtering and randomization.}
We leave this as future work. 


\subsection{More Evaluations}
In \cref{sec:4} we present the feasibility and preliminary results of two proposed attack methodologies. They can be further extended in the following ways. (1) We can optimize the attack designs. For instance, for the obstacle position altering attack, we can test more common objects to generate the fake points. For the location altering attack, we can try to reduce the number of injected obstacles and patches when increasing the similarity. (2) We give some examples to show the success of the attacks. More quantitative evaluations and comparisons can better demonstrate their effectiveness. (3) We evaluate the attacks with the dataset and simulator. Physical experiments will make the results more convincing. In the future, we will improve the attacks from these aspects, and also realize other attack vectors analyzed in \cref{sec:3-analysis}.

\section{Conclusion}
In this paper, we first systematize the knowledge of sensor spoofing attacks against RV systems. Then an action flow model is introduced to describe existing attacks and predict new attack vectors. Our model and analysis can benefit RV researchers and practitioners in understanding the \revision{unexplored sensor spoofing threats}, and inspecting their designs. We also propose two new attack methodologies against the trajectory tracking and loop closure detection. We expect these methodologies can inspire researchers to improve our identified attack vectors, and design the corresponding defenses. 


\section*{Acknowledgements}
We sincerely thank our shepherd, Dr. Tom Chothia, and the anonymous reviewers for their valuable comments on this paper. This work is supported under the RIE2020 Industry Alignment Fund-Industry Collaboration Projects (IAF-ICP) Funding Initiative, as well as cash and in-kind contributions from the industry partner(s). It is also supported in part by NTU-DESAY SV Research Program under Grant 2018-0980 and Singapore Ministry of Education (MOE) AcRF Tier 1 RG108/19 (S).

\clearpage

\bibliographystyle{IEEEtran}
\bibliography{references}

\appendices
\clearpage
\appendices


\setlength{\tabcolsep}{5pt}
\begin{table*}[b!]
\caption{All the possible action flows that can be spoofed and the corresponding works.}
\vspace{-10pt}
\begin{adjustwidth}{0.2cm}{0.2cm}
\begin{tablenotes}[para,flushleft]
\footnotesize
    \centering 
    \textbf{A1}: Object detection \enspace \textbf{A2}: Segmentation \enspace \textbf{A3}: Localization/SLAM \enspace \textbf{A4}: Speech recognition \enspace \textbf{A5}: Distance detection \enspace \textbf{B}: Object tracking \\
    \textbf{C1}: Environment prediction \enspace \textbf{C2}: Goal planning \enspace \textbf{D1}: Path routing \enspace \textbf{D2}: Parking/Landing planning \enspace \textbf{E}: Motion planning \enspace \textbf{F}: Motion controller
\end{tablenotes}
\end{adjustwidth}
\vspace{2pt}
\centering
\arrayrulecolor{black}
\resizebox{1.0\linewidth}{!}{
\begin{tabular}{|c|c|cccccc|ccccc|c|c|c|c|c|}
\rowcolor{gray!40}
\hline
{\cellcolor{gray!40}} & & \multicolumn{6}{c|}{\textbf{Perception}} & \multicolumn{5}{c|}{\textbf{Planning}} & \multicolumn{1}{c|}{\textbf{Control}} & \multirow{2}{*}{\textbf{\begin{tabular}[c]{@{}c@{}}Attack\\\textcolor{gray!40}{X} \ Path\ \textcolor{gray!40}{X}\end{tabular}}} & {\cellcolor{gray!40}} & \multirow{2}{*}{\textbf{\begin{tabular}[c]{@{}c@{}}Action\\\quad\ Flow \quad\ \end{tabular}}} & \multirow{2}{*}{\textbf{\begin{tabular}[c]{@{}c@{}}Flow\\ \quad Pattern \quad \end{tabular}}} \\
\hhline{|>{\arrayrulecolor{gray!40}}-->{\arrayrulecolor{black}}------------>{\arrayrulecolor{gray!40}}-->{\arrayrulecolor{black}}|}
\multirow{-2}{*}{\cellcolor{gray!40}\textbf{Type}} & \multirow{-2}{*}{\cellcolor{gray!40}\textbf{Victim Sensor}} & \multicolumn{1}{c}{\cellcolor{gray!40}A1} & \multicolumn{1}{c}{\cellcolor{gray!40}A2} & \multicolumn{1}{c}{\cellcolor{gray!40}A3} & \multicolumn{1}{c}{\cellcolor{gray!40}A4} & \multicolumn{1}{c}{\cellcolor{gray!40}A5} & \multicolumn{1}{|c|}{\cellcolor{gray!40}B} & \multicolumn{1}{c}{\cellcolor{gray!40}C1} & \multicolumn{1}{c}{\cellcolor{gray!40}C2} & \multicolumn{1}{c}{\cellcolor{gray!40}D1} & \multicolumn{1}{c}{\cellcolor{gray!40}D2} & \multicolumn{1}{|c|}{\cellcolor{gray!40}E} & \multicolumn{1}{c|}{\cellcolor{gray!40}F} & & \multirow{-2}{*}{\cellcolor{gray!40}\textbf{Paper}} & & \\ 
\hline
\multirow{34}{*}{\rotatebox{90}{\large{Single-Round Action Flow}}} & \multirow{2}{*}{WWM Radar}
 & \fullcirc & & & & & \multicolumn{1}{|c|}{} & & & & & \multicolumn{1}{|c|}{\fullcirc} & \fullcirc & \revision{AtkPath4} & \cite{spoof-ultrosonic-defcon16,spoof-ultrosonic-ashes19,spoof-radar-tifs21} & \revision{ActFlow1} & \cellcolor{red!40}{\revision{FlowPtn1}} \\
 \hhline{~~*{16}{-|}}
 & & \fullcirc & & & & & \multicolumn{1}{|c|}{\fullcirc} & \fullcirc & & & & \multicolumn{1}{|c|}{\fullcirc} & \fullcirc & \revision{AtkPath5} & \cite{spoof-radar-tifs21} & \revision{ActFlow2} & \cellcolor{blue!40}{\revision{FlowPtn2}} \\ 
\hhline{~*{17}{-|}}
& \multirow{7}{*}{LiDAR}
 & \fullcirc & & & & & \multicolumn{1}{|c|}{} & & & & & \multicolumn{1}{|c|}{\fullcirc} & \fullcirc & \revision{AtkPath4} & \cite{spoof-lidar-bhe15,spoof-lidar-shin,jin2022pla,spoof-lidar-ccs19,spoof-lidar-us20,spoof-lidar-camera-sp21,spoof-lidar-ccs21}  & \revision{ActFlow3} & \cellcolor{red!40}{\revision{FlowPtn1}} \\
 \hhline{~~*{16}{-|}}
 & & \fullcirc & & & & & \multicolumn{1}{|c|}{\fullcirc} & \fullcirc & & & & \multicolumn{1}{|c|}{} & \fullcirc & \revision{AtkPath5} & - & \revision{ActFlow4} & \cellcolor{blue!40}{\revision{FlowPtn2}} \\ 
 \hhline{~~*{16}{-|}}
 & & & & \fullcirc & & & \multicolumn{1}{|c|}{} & \fullcirc & & & & \multicolumn{1}{|c|}{\fullcirc} & \fullcirc & \revision{AtkPath3} & - & \revision{ActFlow5} & \cellcolor{ForestGreen!40}{\revision{FlowPtn5}} \\
 \hhline{~~*{16}{-|}}
 & & & & \fullcirc & & & \multicolumn{1}{|c|}{} & & & \fullcirc & & \multicolumn{1}{|c|}{\fullcirc} & \fullcirc & \revision{AtkPath3} & - & \revision{ActFlow6} & \cellcolor{orange!40}{\revision{FlowPtn6}} \\
 \hhline{~~*{16}{-|}}
 & & & & \fullcirc & & & \multicolumn{1}{|c|}{} & & & & \fullcirc & \multicolumn{1}{|c|}{\fullcirc} & \fullcirc & \revision{AtkPath2} & - & \revision{ActFlow7} & \cellcolor{purple!40}{\revision{FlowPtn8}} \\
 \hhline{~~*{16}{-|}}
 & & & & \fullcirc & & & \multicolumn{1}{|c|}{} & & & & & \multicolumn{1}{|c|}{\fullcirc} & \fullcirc & \revision{AtkPath3} & - & \revision{ActFlow8} & \cellcolor{green!40}{\revision{FlowPtn4}} \\
 \hhline{~~*{16}{-|}}
 & & & & \fullcirc & & & \multicolumn{1}{|c|}{} & & & & & \multicolumn{1}{|c|}{} & \fullcirc & \revision{AtkPath1} & - & \revision{ActFlow9} & \cellcolor{yellow!40}{\revision{FlowPtn7}} \\ 
\hhline{~*{17}{-|}}
& \multirow{8}{*}{Camera}
 & \fullcirc & & & & & \multicolumn{1}{|c|}{} & & & & & \multicolumn{1}{|c|}{\fullcirc} & \fullcirc & \revision{AtkPath4} & \cite{spoof-camera-stick-cvpr18,spoof-camera-stick-woot18,spoof-camera-stick-ccs19,spoof-camera-raid,yan2022rolling,spoof-camera-light-ccs20,spoof-camera-light-iacr20,spoof-camera-light-cvpr21,spoof-camera-light-ccs21,spoof-camera-light-us21,kong2020physgan,wang2021dual,cheng2022physical,huang2020universal,wu2020making,xu2020adversarial,spoof-camera-uav-us22} & \revision{ActFlow10} & \cellcolor{red!40}{\revision{FlowPtn1}} \\
 \hhline{~~*{16}{-|}}
 & & \fullcirc & & & & & \multicolumn{1}{|c|}{\fullcirc} & \fullcirc & & & & \multicolumn{1}{|c|}{\fullcirc} & \fullcirc & \revision{AtkPath5} & \cite{spoof-camera-stick-iclr20} & \revision{ActFlow11} & \cellcolor{blue!40}{\revision{FlowPtn2}} \\ 
 \hhline{~~*{16}{-|}}
 & & & \fullcirc & & & & \multicolumn{1}{|c|}{} & \fullcirc & & & & \multicolumn{1}{|c|}{\fullcirc }& \fullcirc & \revision{AtkPath4} & \cite{spoof-camera-light-iacr20,spoof-camera-stick-us21,spoof-camera-stick-us21-cq,spoof-camera-light-ccs20} & \revision{ActFlow12} & \cellcolor{Cyan!40}{\revision{FlowPtn3}} \\ 
 \hhline{~~*{16}{-|}}
 & & & & \fullcirc & & & \multicolumn{1}{|c|}{} & \fullcirc & & & & \multicolumn{1}{|c|}{\fullcirc} & \fullcirc & \revision{AtkPath3} & - & \revision{ActFlow13} & \cellcolor{ForestGreen!40}{\revision{FlowPtn5}} \\
 \hhline{~~*{16}{-|}}
 & & & & \fullcirc & & & \multicolumn{1}{|c|}{} & & & \fullcirc & & \multicolumn{1}{|c|}{\fullcirc} & \fullcirc & \revision{AtkPath3} & - & \revision{ActFlow14} & \cellcolor{orange!40}{\revision{FlowPtn6}} \\
 \hhline{~~*{16}{-|}}
 & & & & \fullcirc & & & \multicolumn{1}{|c|}{} & & & & \fullcirc & \multicolumn{1}{|c|}{\fullcirc} & \fullcirc & \revision{AtkPath2} & - & \revision{ActFlow15} & \cellcolor{purple!40}{\revision{FlowPtn8}} \\
 \hhline{~~*{16}{-|}}
 & & & & \fullcirc & & & \multicolumn{1}{|c|}{} & & & & & \multicolumn{1}{|c|}{\fullcirc} & \fullcirc & \revision{AtkPath3} & - & \revision{ActFlow16} & \cellcolor{green!40}{\revision{FlowPtn4}} \\
 \hhline{~~*{16}{-|}}
 & & & & \fullcirc & & & \multicolumn{1}{|c|}{} & & & & & \multicolumn{1}{|c|}{} & \fullcirc & \revision{AtkPath1} & \cite{spoof-camera-light-woot16} & \revision{ActFlow17} & \cellcolor{yellow!40}{\revision{FlowPtn7}} \\ 
\hhline{~*{17}{-|}}
& \multirow{5}{*}{GPS}
 & & & \fullcirc & & & \multicolumn{1}{|c|}{} & \fullcirc & & & & \multicolumn{1}{|c|}{\fullcirc} & \fullcirc & \revision{AtkPath3} & - & \revision{ActFlow18} &  \cellcolor{ForestGreen!40}{\revision{FlowPtn5}}\\
 \hhline{~~*{16}{-|}}
 & & & & \fullcirc & & & \multicolumn{1}{|c|}{} & & & \fullcirc & & \multicolumn{1}{|c|}{\fullcirc} & \fullcirc & \revision{AtkPath3} & \cite{spoof-gps-car-us18,spoof-gps-uav-us22} & \revision{ActFlow19} & \cellcolor{orange!40}{\revision{FlowPtn6}} \\
 \hhline{~~*{16}{-|}}
 & & & & \fullcirc & & & \multicolumn{1}{|c|}{} & & & & \fullcirc & \multicolumn{1}{|c|}{\fullcirc} & \fullcirc & \revision{AtkPath2} & \cite{spoof-gps-drone-vlsid18,spoof-gps-drone-netw19-1,spoof-gps-drone-netw19-2} & \revision{ActFlow20} & \cellcolor{purple!40}{\revision{FlowPtn8}} \\
 \hhline{~~*{16}{-|}}
 & & & & \fullcirc & & & \multicolumn{1}{|c|}{} & & & & & \multicolumn{1}{|c|}{\fullcirc} & \fullcirc & \revision{AtkPath3} & \cite{spoof-gps-car-us20,spoof-gps-uav-us22} & \revision{ActFlow21} & \cellcolor{green!40}{\revision{FlowPtn4}} \\
 \hhline{~~*{16}{-|}}
 & & & & \fullcirc & & & \multicolumn{1}{|c|}{} & & & & & \multicolumn{1}{|c|}{} & \fullcirc & \revision{AtkPath1} & - & \revision{ActFlow22} & \cellcolor{yellow!40}{\revision{FlowPtn7}} \\ 
\hhline{~*{17}{-|}}
& \multirow{5}{*}{IMU}
 & & & \fullcirc & & & \multicolumn{1}{|c|}{} & \fullcirc & & & & \multicolumn{1}{|c|}{\fullcirc} & \fullcirc & \revision{AtkPath3} & - & \revision{ActFlow23} & \cellcolor{ForestGreen!40}{\revision{FlowPtn5}} \\
 \hhline{~~*{16}{-|}}
 & & & & \fullcirc & & & \multicolumn{1}{|c|}{} & & & \fullcirc & & \multicolumn{1}{|c|}{\fullcirc} & \fullcirc & \revision{AtkPath3} & - & \revision{ActFlow24} & \cellcolor{orange!40}{\revision{FlowPtn6}} \\
 \hhline{~~*{16}{-|}}
 & & & & \fullcirc & & & \multicolumn{1}{|c|}{} & & & & \fullcirc & \multicolumn{1}{|c|}{\fullcirc} & \fullcirc & \revision{AtkPath2} & - & \revision{ActFlow25} & \cellcolor{purple!40}{\revision{FlowPtn8}} \\
 \hhline{~~*{16}{-|}}
 & & & & \fullcirc & & & \multicolumn{1}{|c|}{} & & & & & \multicolumn{1}{|c|}{\fullcirc} & \fullcirc & \revision{AtkPath3} & - & \revision{ActFlow26} & \cellcolor{green!40}{\revision{FlowPtn4}} \\
 \hhline{~~*{16}{-|}}
 & & & & \fullcirc & & & \multicolumn{1}{|c|}{} & & & & & \multicolumn{1}{|c|}{} & \fullcirc & \revision{AtkPath1} & \cite{spoof-imu-us15,spoof-imu-eurosp17,spoof-imu-us18} & \revision{ActFlow27} & \cellcolor{yellow!40}{\revision{FlowPtn7}} \\ 
\hhline{~*{17}{-|}}
& \multirow{3}{*}{Microphone}
 & & & & \fullcirc & & \multicolumn{1}{|c|}{} & & \fullcirc & & & \multicolumn{1}{|c|}{} & \fullcirc & \revision{AtkPath5} & \cite{spoof-voice-wc19,spoof-voice-inaudible-tdsc21} & \revision{ActFlow28} & \cellcolor{Magenta!40}{\revision{FlowPtn12}} \\
 \hhline{~~*{16}{-|}}
 & & & & & \fullcirc & & \multicolumn{1}{|c|}{} & & \fullcirc & \fullcirc & & \multicolumn{1}{|c|}{\fullcirc} & \fullcirc & \revision{AtkPath5} & \cite{spoof-voice-wc19,spoof-voice-inaudible-tdsc21} & \revision{ActFlow29} & \cellcolor{Magenta!40}{\revision{FlowPtn11}} \\
 \hhline{~~*{16}{-|}}
 & & & & & \fullcirc & & \multicolumn{1}{|c|}{} & & \fullcirc & & \fullcirc & \multicolumn{1}{|c|}{\fullcirc} & \fullcirc & \revision{AtkPath5} & \cite{spoof-voice-wc19,spoof-voice-inaudible-tdsc21} & \revision{ActFlow30} & \cellcolor{Magenta!40}{\revision{FlowPtn13}} \\
\hhline{~*{17}{-|}}
& \multirow{2}{*}{Ultrasonic Sensor}
 & & & & & \fullcirc & \multicolumn{1}{|c|}{} & & & & \fullcirc & \multicolumn{1}{|c|}{\fullcirc} & \fullcirc & \revision{AtkPath4} & \cite{spoof-ultrosonic-defcon16,spoof-ultrosonic-iotj18} & \revision{ActFlow31} & \cellcolor{GreenYellow!40}{\revision{FlowPtn9}} \\
 \hhline{~~*{16}{-|}}
 & & & & & & \fullcirc & \multicolumn{1}{|c|}{} & & & & & \multicolumn{1}{|c|}{} & \fullcirc & \revision{AtkPath4} & - & \revision{ActFlow32} & \cellcolor{Aquamarine!40}{\revision{FlowPtn10}} \\
\toprule[1.0pt]
\bottomrule[1.0pt]
 \multirow{16}{*}{\rotatebox{90}{\large{Multi-Round Action Flow}}} & IMU \ding{222} Env & & & \fullcirc & & & \multicolumn{1}{|c|}{} & & & & & \multicolumn{1}{|c|}{} & \fullcirc & - & & \ding{242} & \cellcolor{Lavender!60}{} \\
 \cline{3-16}
& Env \ding{222} Camera
 & \fullcirc & & & & & \multicolumn{1}{|c|}{} & & & & & \multicolumn{1}{|c|}{\fullcirc} & \fullcirc & \revision{AtkPath6} & \cite{spoof-imu-sp21} & \revision{ActFlow33} & \cellcolor{Lavender!60}{} \\
 \cline{3-16}
& Env \ding{222} LiDAR
 & \fullcirc & & & & & \multicolumn{1}{|c|}{} & & & & & \multicolumn{1}{|c|}{\fullcirc} & \fullcirc & \revision{AtkPath6} & - & \revision{ActFlow34} & \cellcolor{Lavender!60}{} \\
 \cline{3-16}
& Env \ding{222} Camera
 & \fullcirc & & & & & \multicolumn{1}{|c|}{} & & & & & \multicolumn{1}{|c|}{\fullcirc} & \fullcirc & \revision{AtkPath7} & - & \revision{ActFlow35} & \multirow{-4}{*}{\cellcolor{Lavender!60}{\revision{FlowPtn14}}} \\
\hhline{~*{17}{-|}}
 & Camera \ding{222} Env & & & \fullcirc & & & \multicolumn{1}{|c|}{} & & & & & \multicolumn{1}{|c|}{} & \fullcirc & - & & \ding{242} & \cellcolor{Lavender!60}{} \\
 \cline{3-16}
& Env \ding{222} Camera
 & \fullcirc & & & & & \multicolumn{1}{|c|}{} & & & & & \multicolumn{1}{|c|}{\fullcirc} & \fullcirc & \revision{AtkPath6} & - & \revision{ActFlow36} & \cellcolor{Lavender!60}{} \\
 \cline{3-16}
& Env \ding{222} LiDAR
 & \fullcirc & & & & & \multicolumn{1}{|c|}{} & & & & & \multicolumn{1}{|c|}{\fullcirc} & \fullcirc & \revision{AtkPath6} & - & \revision{ActFlow37} & \cellcolor{Lavender!60}{} \\
 \cline{3-16}
& Env \ding{222} Camera
 & \fullcirc & & & & & \multicolumn{1}{|c|}{} & & & & & \multicolumn{1}{|c|}{\fullcirc} & \fullcirc & \revision{AtkPath7} & - & \revision{ActFlow38} & \multirow{-4}{*}{\cellcolor{Lavender!60}{\revision{FlowPtn14}}} \\
\hhline{~*{17}{-|}}
 & LiDAR \ding{222} Env & & & \fullcirc & & & \multicolumn{1}{|c|}{} & & & & & \multicolumn{1}{|c|}{} & \fullcirc & - & & \ding{242} & \cellcolor{Lavender!60}{} \\
 \cline{3-16}
& Env \ding{222} Camera
 & \fullcirc & & & & & \multicolumn{1}{|c|}{} & & & & & \multicolumn{1}{|c|}{\fullcirc} & \fullcirc & \revision{AtkPath6} & - & \revision{ActFlow39} & \cellcolor{Lavender!60}{} \\
 \cline{3-16}
& Env \ding{222} LiDAR
 & \fullcirc & & & & & \multicolumn{1}{|c|}{} & & & & & \multicolumn{1}{|c|}{\fullcirc} & \fullcirc & \revision{AtkPath6} & - & \revision{ActFlow40} & \cellcolor{Lavender!60}{} \\
 \cline{3-16}
& Env \ding{222} Camera
 & \fullcirc & & & & & \multicolumn{1}{|c|}{} & & & & & \multicolumn{1}{|c|}{\fullcirc} & \fullcirc & \revision{AtkPath7} & - & \revision{ActFlow41} & \multirow{-4}{*}{\cellcolor{Lavender!60}{\revision{FlowPtn14}}} \\
\hhline{~*{17}{-|}}
 & GPS \ding{222} Env & & & \fullcirc & & & \multicolumn{1}{|c|}{} & & & & & \multicolumn{1}{|c|}{} & \fullcirc & - & & \ding{242} & \cellcolor{Lavender!60}{} \\
 \cline{3-16}
& Env \ding{222} Camera
 & \fullcirc & & & & & \multicolumn{1}{|c|}{} & & & & & \multicolumn{1}{|c|}{\fullcirc} & \fullcirc & \revision{AtkPath6} & - & \revision{ActFlow42} & \cellcolor{Lavender!60}{} \\
 \cline{3-16}
& Env \ding{222} LiDAR
 & \fullcirc & & & & & \multicolumn{1}{|c|}{} & & & & & \multicolumn{1}{|c|}{\fullcirc} & \fullcirc & \revision{AtkPath6} & \cite{spoof-gps-car-iccv21} & \revision{ActFlow43} & \cellcolor{Lavender!60}{} \\
 \cline{3-16}
& Env \ding{222} Camera
 & \fullcirc & & & & & \multicolumn{1}{|c|}{} & & & & & \multicolumn{1}{|c|}{\fullcirc} & \fullcirc & \revision{AtkPath7} & \cite{spoof-gps-car-autosec21} & \revision{ActFlow44} & \multirow{-4}{*}{\cellcolor{Lavender!60}{\revision{FlowPtn14}}} \\
\hline
\end{tabular}}
\label{table:action_flow}
\end{table*}

\appendices

\end{document}